\documentclass{aastex6}
\usepackage{graphicx}

\fullcollaborationName{The Friends of AASTeX Collaboration}

\begin{document}


\title{Circumbinary Disks of the Protostellar Binary Systems in the L1551 Region}

%% Use \author, \affil, plus the \and command to format author and affiliation 
%% information.  If done correctly the peer review system will be able to
%% automatically put the author and affiliation information from the manuscript
%% and save the corresponding author the trouble of entering it by hand.
%%
%% The \affil should be used to document primary affiliations and the
%% \altaffil should be used for secondary affiliations, titles, or email.

%% Authors with the same affiliation can be grouped in a single

\author{Shigehisa Takakuwa\altaffilmark{1,2},
Kazuya Saigo\altaffilmark{3}, Tomoaki Matsumoto\altaffilmark{4},
Masao Saito\altaffilmark{5},
Jeremy Lim\altaffilmark{6}, Hsi-Wei Yen\altaffilmark{2},
Nagayoshi Ohashi\altaffilmark{2},
% Munetake Momose\altaffilmark{7},
Paul T. P. Ho\altaffilmark{2,7},
\&
Leslie W. Looney\altaffilmark{8}}
\altaffiltext{1}{Department of Physics and Astronomy, Graduate School of Science and Engineering,
Kagoshima University, 1-21-35 Korimoto, Kagoshima, Kagoshima 890-0065, Japan;
takakuwa@sci.kagoshima-u.ac.jp}
\altaffiltext{2}{Academia Sinica Institute of Astronomy and Astrophysics,
11F of Astro-Math Bldg, 1, Sec. 4, Roosevelt Rd, Taipei 10617, Taiwan}
\altaffiltext{3}{ALMA Project Office, National Astronomical Observatory of Japan, Osawa 2-21-1,
Mitaka, Tokyo 181-8588, Japan}
\altaffiltext{4}{Faculty of Humanity and Environment, Hosei University, Chiyoda-ku, Tokyo 102-8160}
\altaffiltext{5}{TMT-J Project Office, National Astronomical Observatory of Japan, Osawa 2-21-1,
Mitaka, Tokyo 181-8588, Japan}
\altaffiltext{6}{Department of Physics, University of Hong Kong, Pokfulam Road, Hong Kong}
% \altaffiltext{}{Center for Frontier Science, Chiba University, Inage-ku, Chiba 263-8522, Japan}
% \altaffiltext{7}{College of Science, Ibaraki University, Bunkyo 2-1-1, Mito, Ibaraki, 310-8512, Japan}
\altaffiltext{7}{East Asian Observatory, 660 N. A'ohoku Place, University Park, Hilo, Hawaii 96720, U.S.A.}
\altaffiltext{8}{Department of Astronomy, University of Illinois, Urbana, IL, U.S.A.}


\begin{abstract}
We report ALMA Cycle 4 observations of the Class I binary
protostellar system L1551 IRS 5 in the 0.9-mm continuum emission, 
C$^{18}$O ($J$=3--2), OCS ($J$=28--27), and four other Band 7 lines.
At $\sim$0$\farcs$07 (= 10 au) resolution in the 0.9 mm emission,
two circumstellar disks (CSDs) associated with the binary protostars are
separated from the circumbinary disk (CBD). The CBD
is resolved into two spiral arms, one connecting to the CSD
around the northern binary source, Source N, and the other
to Source S. As compared to the CBD in the neighboring
protobinary system L1551 NE, the CBD in L1551 IRS 5 is
more compact ($r \sim$150 au) and the $m$=1 mode of the spirals found
in L1551 NE is less obvious in L1551 IRS 5.
Furthermore, the dust and molecular-line brightness temperatures of CSDs and CBD
reach $>$260 K and $>$100 K, respectively, in L1551 IRS 5, much hotter than those in L1551 NE.
The gas motions in the spiral arms are characterized by rotation
and expansion. Furthermore, the transitions from the CBD to the CSD rotations at around
the L2 and L3 Lagrangian points
and gas motions around the L1 point are identified.
Our numerical simulations reproduce the observed two spiral arms and
expanding gas motion as a result of
gravitational torques from the binary, transitions from the CBD to the CSD rotations,
and the gas motion around the L1 point.
The higher temperature in L1551 IRS 5 likely reflects the inferred FU-Ori event.
% rotation in the CBD becomes faster than the local Keplerian
% rotation in the close vicinity of the protobinary, and the CBD
% exhibits expanding gas motion.
%
% as observed in the C$^{18}$O and OCS emission.
% A gas component located between the binary
% exhibits a north (blue) to south (red) velocity gradient.
% In addition, compact, high-velocity blueshifted emission located
% to the north of Source N is identified.
% The spiral features, fast rotation and expansion in the CBD
% are reproduced with our numerical simulations,
% where gravitational torques from the binary system impart angular momenta
% to the spiral arms and cause expanding gas motion.
%
% The velocity gradient in the gas component inbetween the binary likely reflects the gas flow
% between the binary sources, which is also reproduced with our numerical simulation.
% The origin of the high-velocity blueshifted gas is not clear.
\end{abstract}




\section{Introduction} \label{sec:intro}

Binary and multiple stellar systems are ubiquitous in solar-type stars
\cite{du91b,rag10}.
Recent millimeter interferometric survey observations show that binaries
are more frequent among protostellar sources ($\gtrsim$50$\%$) than main-sequence binaries
\cite{mur13,che13,rei14,tob16}.
Protostellar binary systems have been identified as two closely located dust-continuum sources,
which are considered as circumstellar disks (hereafter CSDs)
surrounding each protostar \cite{loo97,rod98,loo00,lim06,mau10,tob15,tob16,lim16a,lim16b}.
Other disks surrounding both of the CSDs, circumbinary disks (CBDs), have also
been identified \cite{tak04,tak12,tob13,cho14,tan14,tan16,dut14,dut16},
which are considered to be mass reservoirs to the CSDs and protostellar binaries.
Observational studies of the structures and gas motions of those CBDs and their comparisons
among different protobinary systems are crucial
to understand how protostellar binaries grow and how their final masses
and mass ratios are determined.

The L1551 region \cite{lyn62} is a low-mass star-forming region located
to the south of the Taurus molecular cloud complex. The distance to
the L1551 region has been adopted to be 140 pc \cite{eli78}, and
a recent estimate of the distance is 147$\pm$5 pc \cite{con18}\footnote{For direct
comparison with our previous publications,
throughout this paper we adopt the distance $d$ = 140 pc.}.
It contains a famous Class I-II source with planet formation,
HL Tau \cite{alma15}, and two Class I protostellar binaries of L1551 IRS 5 and NE
and a Class II binary XZ Tau \cite{hay09}.
The two Class I protostellar binaries of L1551 IRS 5 and NE are located closely at
a projected distance of $\sim$2$\farcm$5 ($\sim$0.10 pc), and L1551 NE is located
northeast of L1551 IRS 5.
We have conducted ALMA Cycles 0 and 2 observations of
L1551 NE \cite[$T_{bol}$ = 91 K and $L_{bol}$ = 4.2 $L_{\odot}$;][]{fro05}
\cite{tak14,tak17}. L1551 NE
comprises the southeastern source named ``Source A" and the northwestern source
``Source B" \cite{rei00,rei02,lim16b}, where their projected and de-projected
separations are estimated to be $\sim$70 au and $\sim$145 au, respectively \cite{tak14}.
The total binary mass and the mass ratio ($\equiv q$) have been estimated to be $\sim$0.8 $M_{\odot}$
and $q\sim$0.2, respectively \cite{tak12}.
Our ALMA observations in the 0.9-mm dust-continuum emission at $\sim$0$\farcs$2 resolution
have identified two arm-like features in the $r \sim$300 au scale CBD,
as well as the CSDs associated with the individual binary stars.
The southern arm connects to the CSD around Source B, and furthermore,
the distribution of the material in the arms are skewed to the west,
suggesting the presence of the $m = 1$ mode. The gas motions in the CBD
as observed in the C$^{18}$O (3--2) emission exhibit expansion in the arms
and infall in the inter-arm regions.
Our hydrodynamic simulation shows that the gravitational torques from the
non-axisymmetric potential of the binary induce both infall
and outward gas motions with slower and faster rotations than the Keplerian rotations,
respectively, and create the spiral density patterns \cite{mat19}.
The observed spiral features with the $m = 1$ mode and the infall and expanding gas motions
in the CBDs of L1551 NE are well reproduced with our hydrodynamic simulation.

A next important step is to study a protostellar binary system
with different physical properties, such as the binary mass ratio,
separation, and temperature. In this paper,
we report ALMA Cycle 4 Band 7 observations of the protostellar binary system
in the L1551 region, L1551 IRS 5.
L1551 IRS 5 is the brightest Class I protostellar source
in the entire Taurus star-forming region
\cite[$L_{bol}$ = 22 $L_{\odot}$ and $T_{bol}$ = 92 K;][]{fro05}.
The binary protostellar system L1551 IRS 5 consists of two compact sources
in the centimeter and millimeter continuum emission along the north-south
direction with the projected and de-projected separations of $\sim$50 au and $\sim$54 au,
respectively.
% (Source N and Source S).
% The binary protostars exhibit proper motions
% originated from the orbital motions of the binary \cite{rod98,rod03a,lim06,lim16a,vil17}.
Both of the binary sources (Source N and Source S)
drive 3.5 cm radio jets at similar intensities along the southwest to
the northeast direction \cite{rod03b}, and HH objects and high-velocity ($\sim$100 km s$^{-1}$)
wind-angle winds are also associated \cite{pyo05,hay09}.
% CO observations of L1551 IRS 5 have revealed a parsec-scale molecular outflow
% with the southwestward blueshifted and northeastward redshifted lobes \cite{sne80,uch87,sto06,mor06}.
Interferometric observations of L1551 IRS 5 in (sub)millimeter molecular lines
have found a $\sim$2500 au scale, rotating and infalling protostellar envelope
elongated along the northwest to southeast direction
% $i.e.$, perpendicular to the associated jets / outflow axis),
and a Keplerian rotating CBD with the radius of $\sim$70 au
embedded in the infalling envelope \cite{oha96,sai96,mom98,tak04,cho14}.
% and the position and inclination angles of the protostellar envelope
% are estimated to be $\sim$160$\degr$ and $\sim$60$\degr$, respectively
% \cite{oha96,sai96,mom98}. A Keplerian rotating CBD with the radius of $\sim$70 au,
% embedded in the infalling envelope, has also been identified, and 
The total binary mass and the mass ratio have been estimated to be $\sim$0.5 $M_{\odot}$
and $q\sim$1, respectively \cite{rod03a,tak04,lim06,cho14,lim16a}.
% from the Keplerian rotation the total binary mass is estimated to be $\sim$0.5 $M_{\odot}$
% \cite{tak04,cho14}. In contrast to the protostellar binary L1551 NE, the binary
% mass ratio in L1551 IRS 5 is likely close to unity, judging from the mass of the CSDs
% as measured from the 7-mm observations \cite{lim06,lim16a} as well as the dynamical
% center of the Keplerian disk \cite{cho14}.
The latest ALMA imaging of L1551 IRS 5 in the 1.3-mm dust-continuum emission
at an angular resolution of $\sim$0$\farcs$14 has
clearly separated the two CSDs and CBD, and resolved the CBD into a $r \sim$100 au ringlike feature
\cite{cru19}.
From the presence of the first overtone CO absorption bands at 2.3 $\micron$
and the association of the HH objects, L1551 IRS 5 has been classified
as one of the FU Orionis objects \cite{her77,har85,rei97,vor06,qua07,con18}, suggesting a warm environment in the
protobinary system.

The structure of this paper is as follows.
In Section 2 we describe our ALMA observations, data reduction, and imaging.
The results of the 0.9-mm dust-continuum and molecular-line emission
are presented in Section 3, in comparison with our previous results of L1551 NE.
In Section 4 we introduce our numerical model and present detailed physical interpretations
of the observed features.
Section 5 summarizes our main results and discussion.

% \section{Methods} \label{sec:met}
\section{ALMA Observations} \label{sec:obs}

ALMA Cycle 4 observations of L1551 IRS 5 at Band 7 were conducted on 2017 July 27.
The 0.9-mm dust-continuum emission and six submillimeter molecular lines listed
in Table \ref{obslines} were observed simultaneously. The total on-source
integration time was 36 minutes. Table \ref{obs} summarizes the observational parameters.
The precipitable water in the atmosphere ranged from $\sim$0.40 mm to $\sim$0.46 mm,
which was excellent for Band 7 observations. The C40-5/(7) configuration was
adopted, and
% the minimum and maximum baseline lengths were 16.7 m and 3.7 km,
% respectively.
% 2017 July 27 --> C40-5/7 ?
% The projected baseline length on the source is from 15.41 m to 3361.68 m.
the projected baseline length on the source was from 15.4 m to 3.4 km.
This minimum projected baseline length indicates that
for a Gaussian emission distribution with
an FWHM of $\sim$9$\farcs$6 ($\sim$ 1300 au),
the peak flux density recovered is 10$\%$ of the peak flux density of the Gaussian \cite{wil94}.
In its Frequency Division Mode (FDM), the correlator was configured to
cover four independent frequency ranges (basebands), and each baseband
has one spectral window (spw). Each spw has a bandwidth of 468.75 MHz
and 3840 spectral channels. Hanning smoothing was applied to the spectral channels,
resulting in a frequency resolution of 244.14 kHz and thus
a velocity resolution of 0.22 km s$^{-1}$ for the C$^{18}$O line.
Spws 0, 1, and 3 cover the $^{13}$CO (3--2), C$^{18}$O (3--2), and CS (7--6) emission,
respectively. Spw 2 includes the OCS (28--27), HC$^{18}$O$^{+}$ (4--3),
and SO (7$_8$--6$_7$) emission.
% Thus, the present ALMA observation is sensitive to the structures of
% the CBD with its outermost radius of $\sim$300 AU ($\sim$2$\farcs$1) around L1551 NE,
% while largely insensitive to its surrounding envelope structures
% with the outermost extent of $>$20000 AU ($\sim$143$\arcsec$) \cite{tak15}.
%
% 0   3840   TOPO  330835.889      -122.070    468750.0 330601.5747        1  XX  YY
% 1   3840   TOPO  329578.423      -122.070    468750.0 329344.1087        2  XX  YY
% 2   3840   TOPO  340410.414       122.070    468750.0 340644.7275        3  XX  YY
% 3   3840   TOPO  342662.651       122.070    468750.0 342896.9652        4  XX  YY
%
% D/l 6.5 arcsec
% theta50 = 2.82705171459424 (arcsec)
% theta10 = 5.18292814342277 (arcsec)

% central freq = 336.107 GHz ~0.891955 mm
% Total Bandwidth = 1.675 GHz

The raw visibility data was calibrated by the ALMA observatory
through the pipeline using the CASA (Common Astronomy Software Applications)
version 5.1.1. The flux, passband, and gain calibrators are quasars listed in
Table \ref{obs} with their estimated flux densities.
The accuracy of the absolute flux calibration of ALMA Band 7 observations is 10$\%$
\footnote{https://almascience.nao.ac.jp/documents-and-tools/cycle8/alma-proposers-guide}.
We also independently
checked the calibrated visibility data and confirmed that the pipeline calibration
was fine. Line-free channels in all the four spectral windows
were used to create the continuum image, which has
a central frequency of 336.11 GHz (= 0.892 mm) and a total bandwidth of 1.675 GHz.
For the continuum imaging, uniform weighting of the visibility data and multi-scale clean
with multiscale=[0,4,20] and cell='0.02arcsec' were adopted, and phase-only
self-calibration was applied. The first iteration of the self-calibration with a solution interval
of 120 seconds achieved a $\sim$20$\%$ reduction of the noise, and after the self-calibration
the spiral features emerged clearly (Figure \ref{fig:cont}). The second iteration of
the phase-only self-calibration with a solution interval of 30 seconds,
or the phase and amplitude self-calibration with the same interval,
did not improve the image quality and the noise, and thus we adopt the
continuum image after the first self-calibration.
The phase-only calibration table
was applied to the line visibility data after the continuum subtraction.  
Natural weighting and multi-scale clean with multiscale=[0,5,25] and cell='0.02arcsec'
were adopted to produce the line images.
The resultant angular resolutions and rms noise levels in the images
are summarized in Table \ref{obs}.
% The achieved line sensitivity measured in the emission-free channels is
% approximately consistent with the theoretically-predicted sensitivity under the best
% weather condition. On the other hand, the continuum noise is a factor of $\sim$5
% worse than the theoretical noise, which could be due to the contamination from
% the extended emission components.

Conversion from the observed intensity $I_{\nu}$ (Jy beam$^{-1}$) to the brightness
temperature $T_B$ (K) is expressed as
\begin{equation}
% T_{B}=\frac{c^2 I_{\nu}}{2 k_{B}\nu^2 \Omega_{\rm beam}},
T_{B}=\frac{c^2 I_{\nu}}{2 k_{B}\nu^2},
\end{equation}
where $\nu$ is the frequency, $c$ speed of light, and $k_B$ the Boltzmann constant.
% and $\Omega_{\rm beam}$ the solid angle of the synthesized beam.
The calculated conversion factors from $I_{\nu}$ to $T_B$ are also listed in Table \ref{obs}.
In this paper results are shown in the unit of $T_B$, to discuss the observed physical conditions.

\begin{deluxetable}{llllll}
\tablecaption{Observed Molecular Lines \label{obslines}}
\tabletypesize{\scriptsize}
\tablewidth{0pt}
% \tablewidth{250pt} % use this in emulateapj
\tablehead{\colhead{Molecule} &\colhead{Transition} &\colhead{Frequency} &\colhead{$E_u$\tablenotemark{a,b}} &\colhead{$A$\tablenotemark{b,c}}        &\colhead{$n_{crit}$\tablenotemark{b,d}}\\
           \colhead{}         &\colhead{}           &\colhead{(GHz)}     &\colhead{(K)}   &\colhead{(s$^{-1}$)} &\colhead{(cm$^{-3}$)}}
\startdata
C$^{18}$O        &$J$=3--2           &329.3305453  &31.6  &2.2$\times$10$^{-6}$  & 3.3$\times$10$^{4}$ \\
$^{13}$CO        &$J$=3--2           &330.587965   &31.7  &2.2$\times$10$^{-6}$  & 3.3$\times$10$^{4}$ \\
OCS              &$J$=28--27         &340.449273   &237.0 &1.2$\times$10$^{-4}$  & 1.6$\times$10$^{6}$ \\
HC$^{18}$O$^{+}$ &$J$=4--3           &340.630700   &40.9  &3.1$\times$10$^{-3}$  & 7.8$\times$10$^{6}$ \\
SO               &$J_N$=7$_8$--6$_7$ &340.714160   &81.2  &5.0$\times$10$^{-4}$  & 7.3$\times$10$^{6}$ \\
CS               &$J$=7--6           &342.882857   &65.8  &8.4$\times$10$^{-4}$  & 1.5$\times$10$^{7}$ \\
\enddata
\tablenotetext{a}{Upper-state energy of the rotational level.}
\tablenotetext{b}{From LAMDA database \cite{sch05}.}
\tablenotetext{c}{Einstein A Coefficient.}
\tablenotetext{d}{Critical density at $T_K$ = 60 K derived from $A/C$, where $C$ denotes the collisional coefficient.}
\end{deluxetable}

\begin{deluxetable}{ll}
\tablecaption{Parameters for the ALMA Cycle 4 Observations of L1551 IRS 5 \label{obs}}
\tabletypesize{\scriptsize}
\tablewidth{0pt}
% \tablewidth{250pt} % use this in emulateapj
\tablehead{\colhead{Parameter} &\colhead{Value}}
\startdata
Observing date & 2017 Jul. 27\\
Number of antennas & 40\\
Field Center & (04$^{h}$31$^{m}$34$\fs$14, 18$\degr$08$\arcmin$05$\farcs$1) (J2000) \\ 
Primary beam HPBW & $\sim$18$\arcsec$ \\
% From CASA image imhead
Central Frequency (Continuum) & 336.107 GHz ($\sim$0.89 mm)\\
Bandwidth (Continuum) & 1.675 GHz \\
Frequency resolution (C$^{18}$O) & 244.14 kHz $\sim$0.22 km s$^{-1}$\\
Synthesized beam HPBW (Continuum; Uniform) & 0$\farcs$0788$\times$0$\farcs$0652 (P.A. = 71.7$\degr$) \\
Synthesized beam HPBW (C$^{18}$O; Natural) & 0$\farcs$116$\times$0$\farcs$114 (P.A. = 22.1$\degr$)  \\
Projected baseline coverage & 15.4 m - 3.4 km \\
Conversion Factor (Continuum) & 1 (Jy beam$^{-1}$) = 2104.2 (K)\\
Conversion Factor (C$^{18}$O) & 1 (Jy beam$^{-1}$) = 852.8 (K)\\
System temperature &$\sim$100 - 350 K \\
rms noise level (Continuum)& 0.32 mJy beam$^{-1}$ = 0.67 K\\
rms noise level (C$^{18}$O)& 5.3 mJy beam$^{-1}$ = 4.5 K\\
Flux calibrator     &J0423-0120 ($\sim$0.66 Jy) \\
Passband calibrator &J0510+180 ($\sim$1.08 Jy)  \\
Gain calibrator &J0440+1437 ($\sim$0.19 Jy), J0431+1731 ($\sim$0.09 Jy) \\
\enddata
\end{deluxetable}

\section{Results} \label{sec:res}
\subsection{0.9-mm Dust-Continuum Emission} \label{subsec:cont}

\begin{figure}[ht!]
\figurenum{1}
\epsscale{1.1}
% \plotone{irs5contbeta.pdf}
% \plotone{irs5contalpha.png}
\plotone{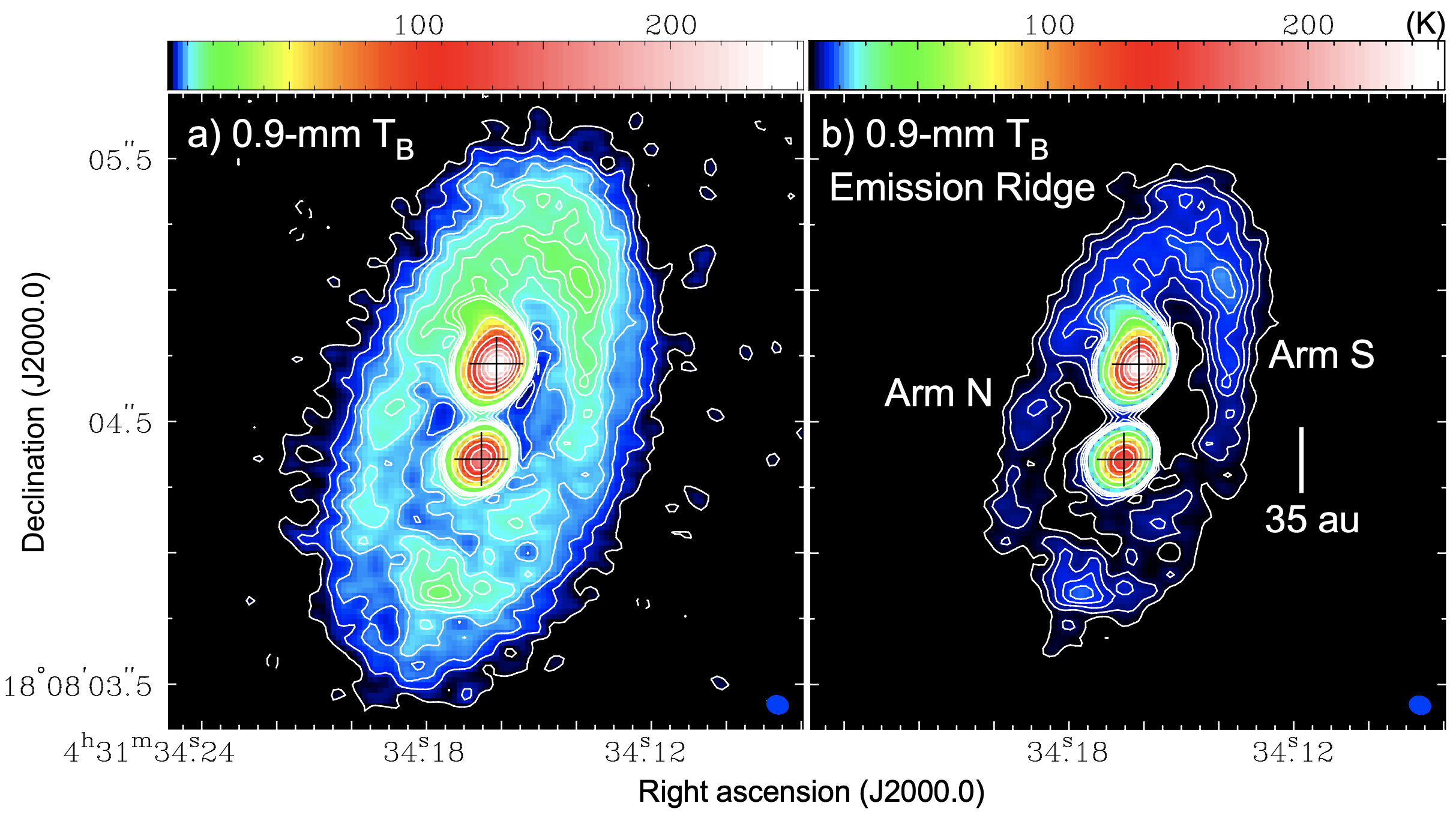}
\caption{The 0.9-mm dust-continuum image of L1551 IRS 5 observed with ALMA.
Upper and lower crosses indicate the centroid positions of
the 2-dimensional Gaussian fittings to the central two dusty components,
which we regard as the positions of Source N and Source S.
A filled ellipse at the bottom-right corner shows the synthesized beam
(0$\farcs$0788$\times$0$\farcs$0652; P.A. = 72$\degr$).
In panel a, contour levels are in steps of 3$\sigma$ until 30$\sigma$,
then 60$\sigma$, 100$\sigma$, and then in steps 50$\sigma$ (1$\sigma$ = 0.67 K).
The color denotes the brightness temperature range from 2.1 to 252.5 K
in log scale.
In panel b, contour levels start from 12$\sigma$, and the subsequent contour
levels are the same as those in panel a. The color scale ranges from 8.1 to 252.5 K.
\label{fig:cont}}
\end{figure}

Figure \ref{fig:cont} presents the ALMA Cycle 4 image of L1551 IRS 5
in the 0.9-mm dust-continuum emission with uniform weighting providing an angular resolution of
% 0$\farcs$0788$\times$0$\farcs$0652 (P.A. = 71.7$\degr$).
0$\farcs$079$\times$0$\farcs$065 (P.A. = 72$\degr$).
In Figure \ref{fig:cont}a the overall extent of the continuum emission
is shown with the lowest contour level of 3$\sigma$, while in
Figure \ref{fig:cont}b the lowest contour level is raised to 12$\sigma$
to show the structure of the emission ridge more clearly.
Two bright components located to the north and south are seen,
which most likely trace CSDs around Source N and Source S, respectively.
The peak brightness temperature toward Source N exceeds $\gtrsim$260 K
and that toward Source S $\gtrsim$160 K.
Previous 7-mm observations of L1551 IRS 5 with JVLA at a
higher angular resolution (0$\farcs$056$\times$0$\farcs$053; P.A. = -1.7$\degr$)
show even higher peak brightness temperatures, $\gtrsim$320 K and $\gtrsim$270 K toward
Source N and Source S, respectively \cite{lim06,lim16a}. These results indicate that
the CSDs around the binary protostars, in particular Source N,
are hot, and it is likely that the dust-continuum emission originated from
those CSDs is optically thick.
From the 2-dimensional Gaussian fittings,
the centroid positions of the 0.9-mm dust-continuum emission are
(04$^h$31$^m$34$\fs$161, +18$\degr$08$\arcmin$04$\farcs$722) toward Source N and
(04$^h$31$^m$34$\fs$165, +18$\degr$08$\arcmin$04$\farcs$359) toward Source S,
which are regarded as the positions of the relevant sources.
The binary protostars in L1551 IRS 5 are known to exhibit
a relative proper motion originated from the binary orbital motion \cite{rod03a,lim06,lim16a}.
To compare the locations of the binary as observed
with ALMA in 2017 to those from the previous measurements
without a possible uncertainty
of the sky proper motion and the absolute astrometry,
the relative position of Source S with respect to Source N
(dRA, dDec) = (0$\farcs$057,-0$\farcs$363) is adopted.
As compared to the relative position derived from
the VLA 7-mm observations of L1551 IRS 5 made in 2012 \cite{lim16a},
the shift from 2012 to 2017 is measured to be
$\sim$(0$\farcs$028,-0$\farcs$007). The $\sim$30 mas shift in RA
over the five years is significant as compared to the
ALMA and JVLA beam sizes over the S/N, and Source S
has moved eastward over the five years. The relative position in 2017
is indeed consistent with the global trend of the orbital motion
starting from 1983 \cite[see Figure 6 of][]{lim16a}.
In particular, in 1983 Source N is located more eastward than Source S
\cite{rod03a}, in 1997 Source N and Source S are
located perfectly north-south \cite{rod98}, and in 2012-2017
Source S is moving eastward with respect to Source N.
These results demonstrate the presence of the clockwise
orbital motion of the binary, which is consistent with the rotational direction
of the CBD \cite{tak04,cho14}.

The beam-deconvolved sizes of the CSDs around Source N and Source S are
measured to be 0$\farcs$175 $\times$ 0$\farcs$123 (25 au $\times$ 17 au) (P.A.=160$\degr$$\pm$5$\degr$)
and 0$\farcs$143 $\times$ 0$\farcs$109 (20 au $\times$ 15 au) (P.A.$\sim$150$\degr$; not well-constrained), respectively.
% 25 au $\times$ 17 au (P.A.=-20$\degr$) and 20 au $\times$ 15 au (P.A.=-33$\degr$),
% respectively.
The JVLA 7-mm observations of L1551 IRS 5 show that
the size of the CSDs around Source N and Source S are 16 au $\times$ 8 au (P.A.=165$\degr$$\pm$3$\degr$)
and 18 au $\times$ 10 au (P.A.=158$\degr$$\pm$5$\degr$), respectively \cite{lim06,lim16a}.
The sizes of the CSDs as seen in the 0.9-mm emission are slightly larger than those in the 7-mm emission,
which could be attributed to the different dust emissivity at the different wavelengths.
The measured CSD position angle of $\sim$160$\degr$ is consistent
with that of the large-scale ($\sim$2500 au) protostellar envelope \cite{mom98},
and that of the CBD derived from our previous SMA observations of L1551 IRS 5 \cite{cho14}.
The de-projected binary separation is estimated to be $\sim$54 au, and the truncation of
the CSDs caused by the tidal interaction of the binary should limit the
CSD diameter to be within one half of the separation \cite{bat00}.
The measured CSD sizes are within this limit.

As well as the two bright CSD components at the center, there is an emission component
surrounding these CSDs, $i.e.,$ the CBD. The outermost extent of the CBD is $\sim$150 au
(Figure \ref{fig:cont}a).
As a whole, the emission elongation of the CBD appears to be
consistent with the position angle of 160$\degr$ as described above, although
it is not straightforward to measure the position angle of the CBD because of its internal structures.
The aspect ratio of the major and minor axes at the 3$\sigma$ level is measured to be $\sim$2,
suggesting an inclination angle of the CBD of $\sim$60$\degr$. This inclination angle is
also consistent with that of the protostellar envelope \cite{mom98} and the JVLA images.
The aspect ratios of the CSDs as seen in the 0.9-mm continuum emission are, however, smaller,
which could be due to the imperfect separation between the CSD and CBD components at 0.9-mm.
Throughout this paper, disk position and inclination angles of 160$\degr$ and 60$\degr$,
and the co-planar configuration of the CBD and CSDs, are assumed.
In the CBD, there is an arm-like feature starting from
a northeastern emission protrusion of the CSD toward Source N (see Figure \ref{fig:cont}b).
This arm curls from east to south.
Hereafter in this paper this arm is called as ``Arm N".
Another arm-like feature starting from
the southwest of Source S and extending to the north is seen (``Arm S").
While this arm appears to merge with the emission protrusion
from the CSD toward Source N, we regard that the starting point of this arm
is from Source S, and that the starting point of Arm N is from Source N.
This interpretation is consistent with the clockwise rotation of the CBD
\cite{tak04,cho14} and the binary orbital motion described above.
With our higher-resolution observations coupled with the self-calibration,
the ringlike CBD previously identified by \citealp[][]{cru19}
has been resolved into two spiral arms.

\begin{figure}[ht!]
\figurenum{2}
\epsscale{0.7}
% \plotone{NEIRS5cont.pdf}
\plotone{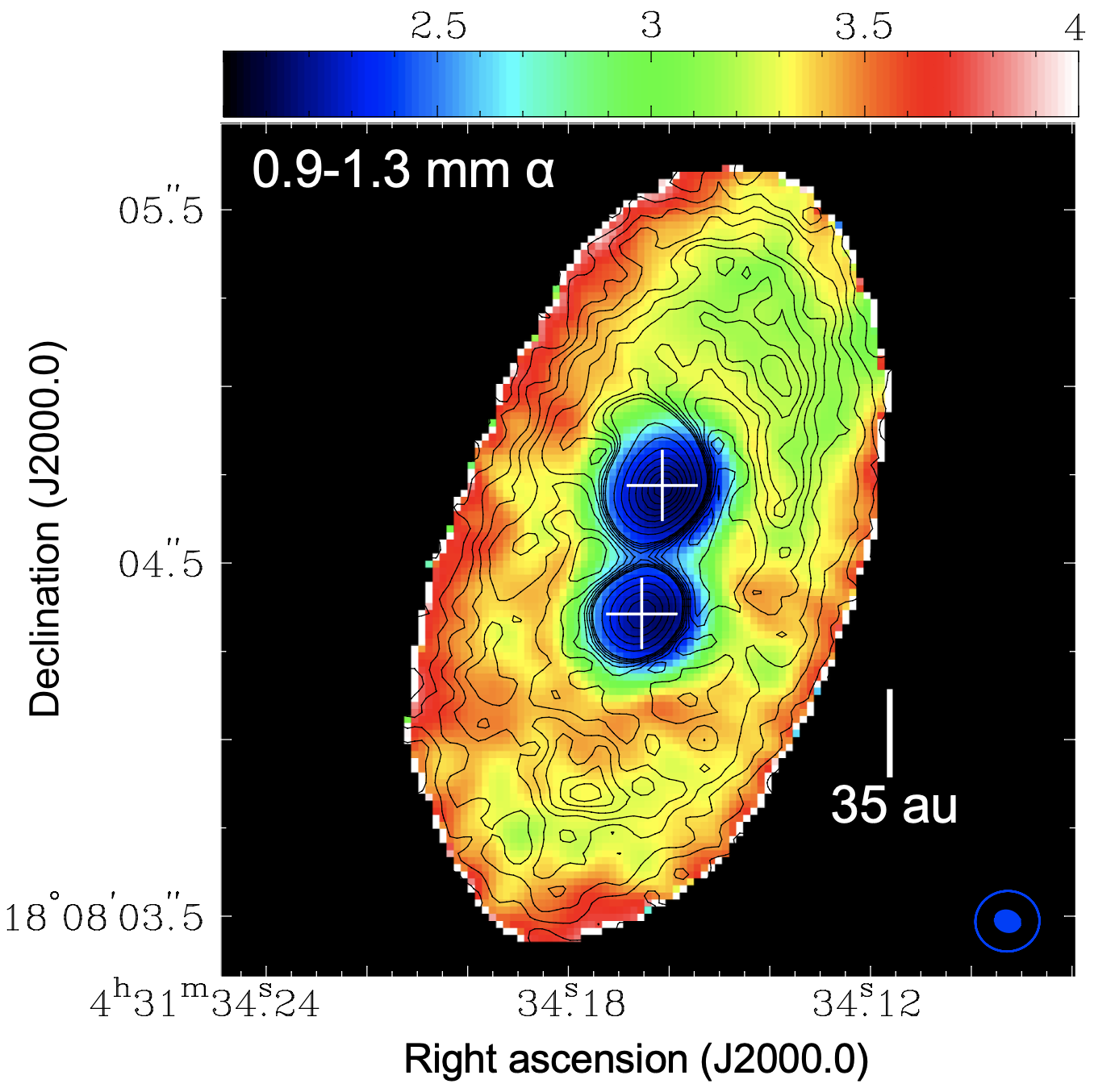}
\caption{Map of the spectral index $\alpha$ deduced from the archival 1.3-mm data
of L1551 IRS 5 and the present 0.9-mm data smoothed to the lower resolution.
Contours denote the 0.9-mm dust-continuum map,
where the contour levels are the same as those of Figure \ref{fig:cont}a. An open ellipse at
the bottom-right corner shows the beam of the $\alpha$ map (0$\farcs$181$\times$0$\farcs$171; P.A. = -83$\degr$),
while the filled ellipse the same as that in Figure \ref{fig:cont}.
\label{fig:alpha}}
\end{figure}

The peak brightness temperatures toward the northwestern and southern peaks in the CBD
are $\sim$20 K and $\sim$18 K, respectively, and the typical brightness temperature
in the CBD is $\lesssim$15 K. As will be discussed in the next subsection,
the peak gas temperature of the CBD is likely as high as $\gtrsim$100 K.
The 0.9-mm dust-continuum emission in the CBD is thus likely optically thin. 
% To further investigate the properties of the dust-continuum emission in L1551 IRS 5,
Figure \ref{fig:alpha} shows the map of the spectral index $\alpha$
of the dust emission deduced from the present 0.9-mm and the published 1.3-mm data \cite{cru19}.
To make this map the 1.3-mm map is corrected for the positional shift originated
from the sky proper motion, and the 0.9-mm map is convolved to have the same beam size
as that of the 1.3-mm map (0$\farcs$18 $\times$ 0$\farcs$17; P.A.=-83$\degr$).
The spectral-index map is made within the pixels where the image intensities at both
wavelengths are above 10$\sigma$.
On the assumption of the 10$\%$ absolute flux uncertainties at both
wavelengths, the uncertainty of $\alpha$ is dominated by the absolute flux uncertainties,
and it is calculated to be $\sim\pm$0.5.
The $\alpha$ value in the CBD is $\sim$3.0 - 3.7, while that in the CSDs is as low as $\sim$2.
The difference of $\alpha$ between the CBD and CSDs, and the difference between
the higher $\alpha$ value ($\gtrsim$3.8) in the outer part of the CBD and the lower value
in the inner parts ($\sim$3.2) are probably real.
These results are consistent with our interpretation that the dust emission toward
the CSDs and CBD are optically thick and thin, respectively.

The mass of the CBD ($\equiv M_{d}$)
can be estimated from the 0.9-mm continuum flux density ($\equiv S_{\nu}$)
with the conventional formula as
\begin{equation}
M_{d}=\frac{S_{\nu}d^2}{\kappa_{\nu} B_{\nu}(T_d)}.
\end{equation}
In the above expression $d$ denotes the distance, $B_{\nu}(T_d)$ the Planck function
for dust at a temperature $T_{d}$, and $\kappa_{\nu}$ the dust opacity per unit gas + dust mass
on the assumption of a gas-to-dust mass ratio of 100.
Assuming $\kappa_{\nu}$ = $\kappa_{\nu_{0}}$($\nu$/$\nu_{0}$)$^{\beta}$,
$\kappa_{\rm 250\mu m}$=0.1 cm$^{2}$ g$^{-1}$ \cite{hil83}, and $\beta$ (= $\alpha$ - 2) =1.0 (Figure \ref{fig:alpha}),
$\kappa_{\rm 0.9mm}$ is calculated to be 0.028 cm$^{2}$ g$^{-1}$.
With this mass opacity and $T_d$ = 30 -- 100 K, the flux density originated from the CBD
($S_{\rm 0.9mm}$ = 1.40 Jy) gives the disk mass of  0.015 -- 0.060 $M_{\odot}$.
If the mass opacity by \citealp[][]{oss94} for grains
with thin ice mantles coagulated at a density of 10$^{6}$ cm$^{-3}$
($\kappa_{\rm 0.9mm}$ = 0.018 cm$^{2}$ g$^{-1}$) is adopted, the mass
estimate of the CBD becomes $\sim$60$\%$ higher.
In either case, the CBD mass is likely much smaller than the inferred
total binary mass of 0.5 $M_{\odot}$ \cite{cho14}, and thus the self-gravity of the CBD
can be ignored compared to the gravitational field of the binary protostars.

\begin{figure}[ht!]
\figurenum{3}
\epsscale{1.2}
% \plotone{NEIRS5cont.pdf}
\plotone{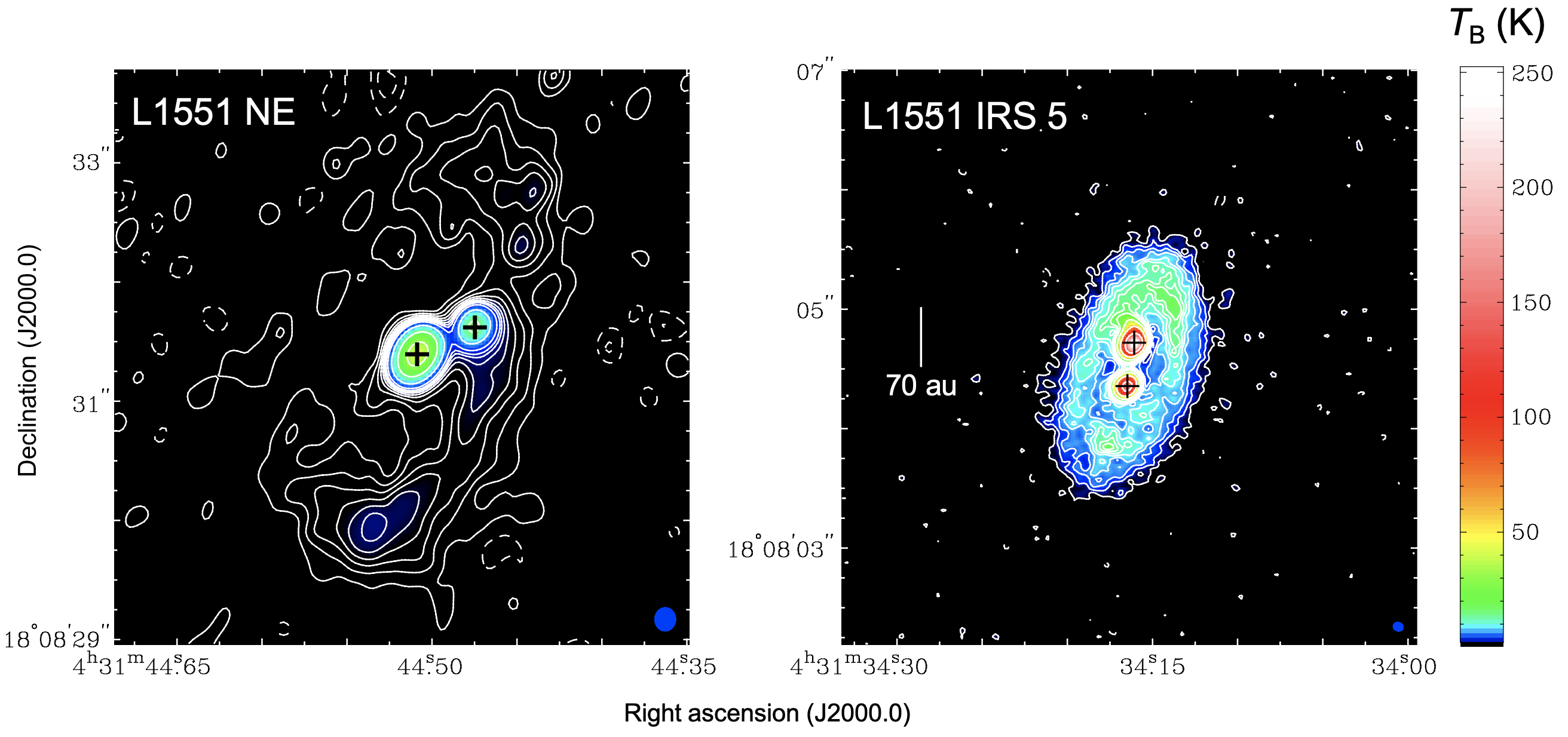}
\caption{Comparison of the 0.9-mm dust-continuum images
in L1551 NE (left) and L1551 IRS 5 (right) at the same linear scale.
Crosses show the positions of the binary protostars. Contour levels are in steps 3$\sigma$
until 24$\sigma$, and then 30$\sigma$, 50$\sigma$, 100$\sigma$, 200$\sigma$, and 400$\sigma$
(1$\sigma$ = 0.162 K in the L1551 NE image and 0.67 K in the L1551 IRS 5 image).
The color denotes the common brightness temperature range from 2.1 to 252.5 K in log scale.
A filled ellipse at the bottom-right corner in each panel shows the relevant synthesized beam,
and the beam size of the L1551 NE image is
0$\farcs$190$\times$0$\farcs$169 (P.A. = -1.9$\degr$).
\label{fig:neirs5}}
\end{figure}

Figure \ref{fig:neirs5} compares the ALMA 0.9-mm dust-continuum images
of L1551 NE \cite[left;][]{tak17} and L1551 IRS 5 (right). While both of the CBDs
exhibit two spiral arms, there are three intriguing
differences between the two CBDs: the size, structure, and the brightness.
The outermost radius of the CBD in L1551 IRS 5 is $\sim$half of that in L1551 NE.
Our Cycle 4 observations of L1551 IRS 5 should be able to recover the emission
components as large as $\sim$10$\arcsec$ (see Section \ref{sec:obs}).
The compact size of the CBD in L1551 IRS 5 ($\sim$2$\arcsec$) is thus not
due to the interferometric filtering effect.
The brightness temperatures of the CSDs
and CBD in L1551 IRS 5 are much higher than those in L1551 NE.
In L1551 NE, the peak brightness temperatures of the CSDs
toward Source A and Source B are 42 K and 18 K, respectively, which are a
factor of $\sim$6 lower than that toward Source N in L1551 IRS 5.
The peak brightness temperature of the CBD in L1551 NE is $\sim$3 K,
a factor of $\sim$6 lower than that in L1551 IRS 5.
Furthermore, the emission distribution in the CBD around L1551 NE
is skewed to the west and exhibits $m$ = 1 mode. Such a mode is much less
clear in the CBD around L1551 IRS 5.
% and the CBD in L1551 IRS 5 consists of two spiral arms with similar intensities.
These differences are likely to
reflect differences of the physical conditions between the two protostellar binaries,
which will be discussed in Section \ref{sec:dis}.

% r=cos i
% dr/di = -sin i
% di=dr / -sin i

\subsection{Molecular-Line Emission} \label{subsec:mol}

\begin{figure}[ht!]
\figurenum{4}
\epsscale{1.0}
% \plotone{irs5mom0sg.pdf}
\plotone{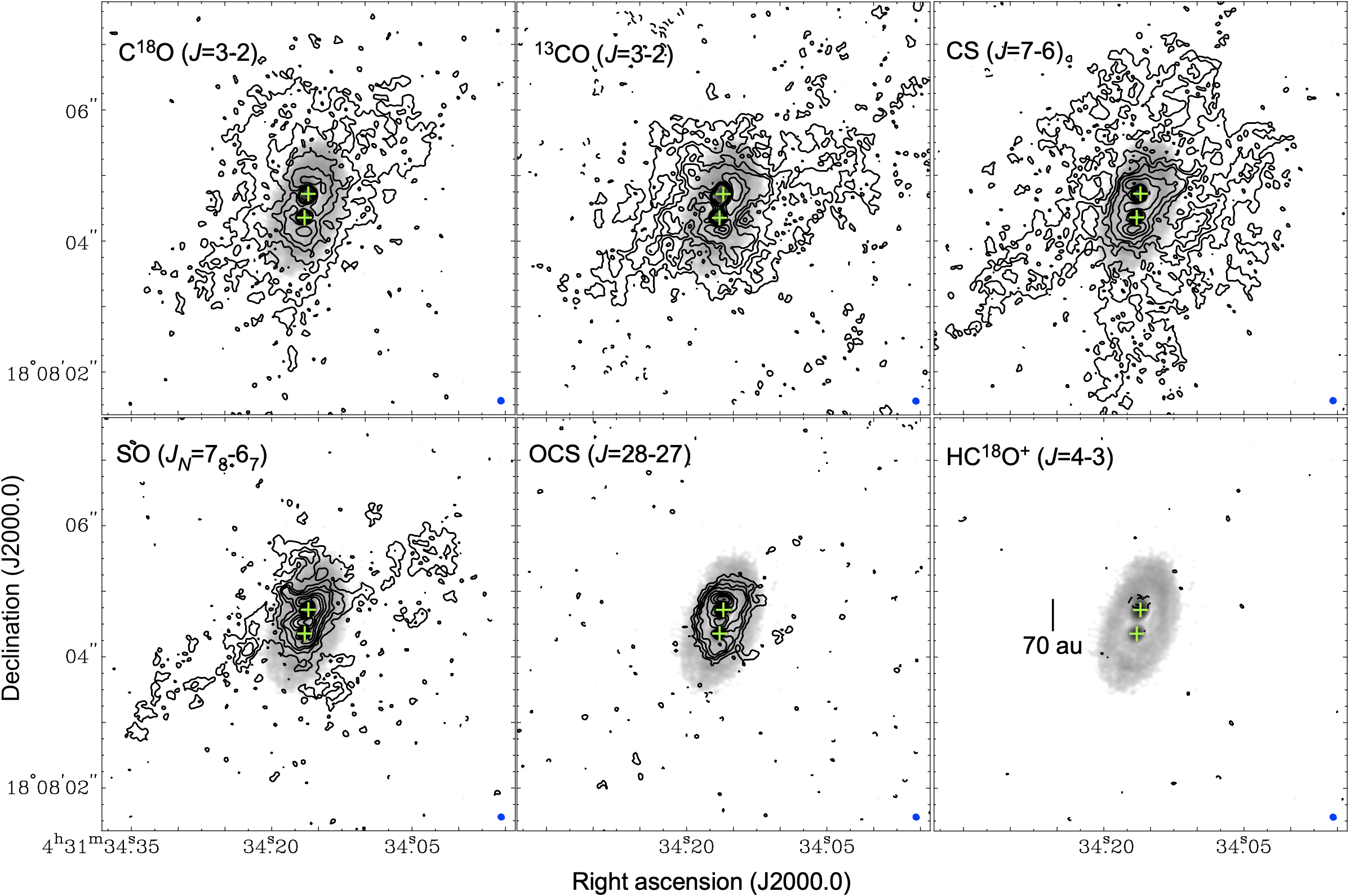}
\caption{Moment 0 maps of the observed molecular lines as labeled (contours),
superimposed on the 0.9-mm continuum image (gray; same as Figure \ref{fig:cont}a), in L1551 IRS 5.
Contour levels are in steps of 5$\sigma$ until 20$\sigma$, and then in steps of 10$\sigma$.
The 1$\sigma$ levels are
5.5 mJy beam$^{-1}$ km s$^{-1}$ (= 4.7 K km s$^{-1}$),
4.8 mJy beam$^{-1}$ km s$^{-1}$ (= 4.4 K km s$^{-1}$),
4.5 mJy beam$^{-1}$ km s$^{-1}$ (= 3.8 K km s$^{-1}$),
4.2 mJy beam$^{-1}$ km s$^{-1}$ (= 3.7 K km s$^{-1}$),
3.7 mJy beam$^{-1}$ km s$^{-1}$ (= 3.3 K km s$^{-1}$),
and
3.4 mJy beam$^{-1}$ km s$^{-1}$ (= 3.0 K km s$^{-1}$)
in the C$^{18}$O, $^{13}$CO, CS, SO, OCS, and HC$^{18}$O$^{+}$ maps, respectively.
Integrated velocity ranges are in the
$V_{\rm LSR}$ = 0.95 -- 10.73 km s$^{-1}$,
1.29 -- 13.80 km s$^{-1}$,
1.25 -- 12.03 km s$^{-1}$,
0.79 -- 14.00 km s$^{-1}$,
1.10 -- 11.64 km s$^{-1}$, and
2.00 -- 10.59 km s$^{-1}$ in the
C$^{18}$O, $^{13}$CO, CS, SO, OCS, and HC$^{18}$O$^{+}$ maps, respectively.
A filled ellipse at the bottom-right corner in each panel shows the relevant synthesized beam.
\label{fig:lines}}
\end{figure}

Figure \ref{fig:lines} presents the moment 0 maps of the observed molecular lines
summarized in Table \ref{obslines} (contours),
superimposed on the 0.9-mm dust-continuum image (gray scale), in L1551 IRS 5.
The C$^{18}$O (3--2) emission appears to trace
the overall structure of the gas component originated from the CBD,
as well as emission extensions toward the northwest and northeast.
In the $^{13}$CO (3--2), CS (7--6), and SO (7$_8$--6$_7$) emission,
there are elongated emission components to
the northwest, northeast, southeast, and the southwest,
as well as the molecular-gas component associated with the CBD.
In L1551 IRS 5, molecular outflows with a wide opening
angle have been observed, and the blueshifted outflow is located to the southwest
and the redshifted outflow to the northeast \cite{sne80,uch87,sto06,mor06}.
The identified emission extensions
are likely to trace the cavity wall of the outflow or the interacting surface
between the protostellar envelope and the outflow.

Figure \ref{fig:lines} also shows that the OCS (28--27) emission, with the
upper-state energy of the rotational level of 237 K, is strongly detected in L1551 IRS 5,
which is not detected in L1551 NE. The emission extent of the OCS emission 
is more compact than that of the dust-continuum emission. These results suggest
that the OCS emission traces the gas components in the inner, hotter part of the
CBD around L1551 IRS 5. On the other hand, the HC$^{18}$O$^{+}$ (4--3) emission
is not detected in L1551 IRS 5 or L1551 NE. Because of the intense 0.9-mm
dust continuum emission toward the CSDs in L1551 IRS 5, all the molecular lines
show negative intensities toward the CSDs.

\begin{figure}[ht!]
\figurenum{5}
\epsscale{1.0}
% \plotone{irs5TBpeaks.pdf}
\plotone{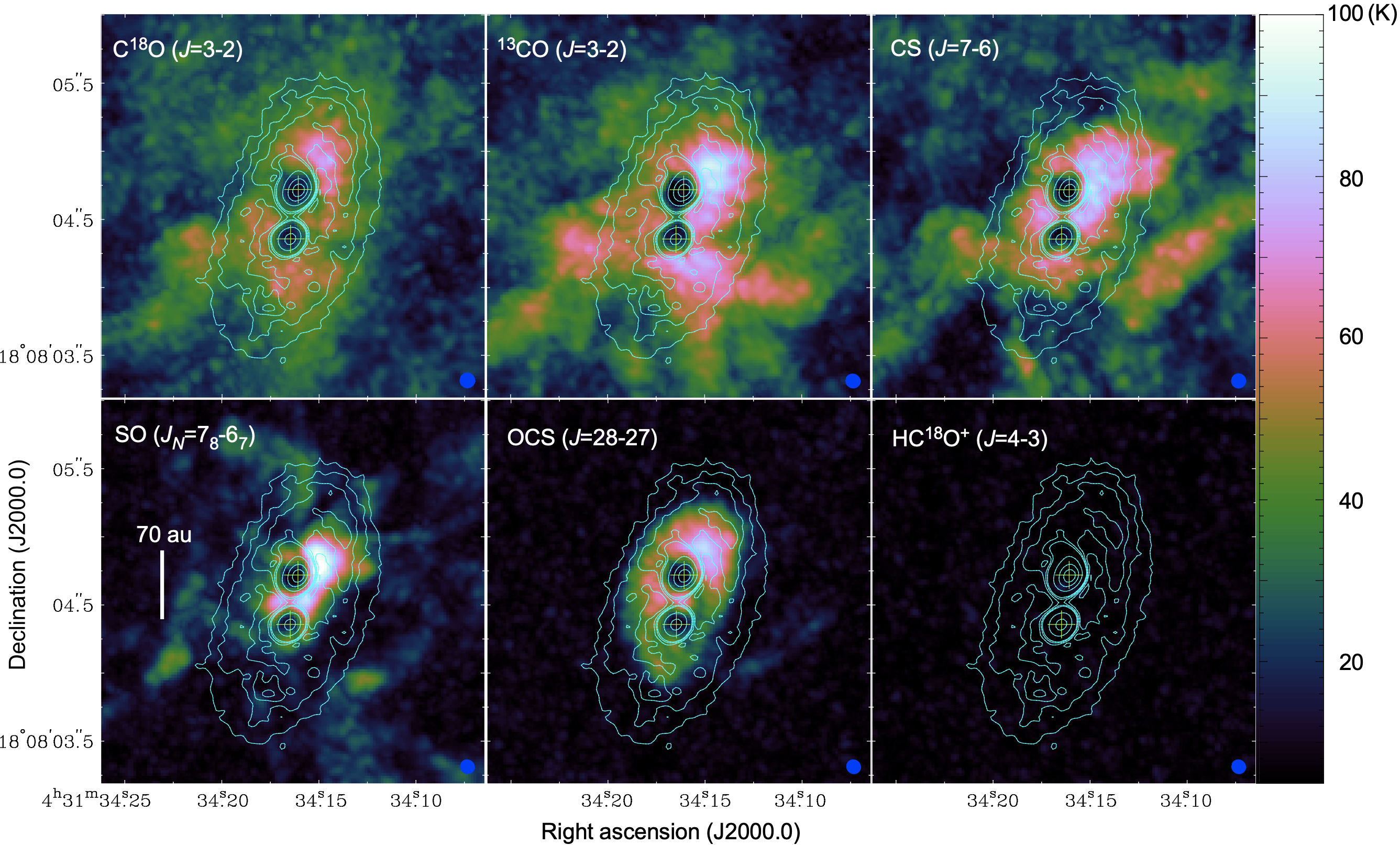}
\caption{Maps of the peak brightness temperatures of
the observed molecular spectra in colors (moment -2 maps),
along with the 0.9-mm dust-continuum image in contours, in L1551 IRS 5.
The color scale ranges from 5 K to 100 K. Contour levels are in steps of 6$\sigma$ until 30$\sigma$,
and then 100$\sigma$, and then in steps of 100$\sigma$ (1$\sigma$ = 0.67 K).
A filled ellipse at the bottom-right corner in each panel shows the relevant synthesized beam.
\label{fig:tb}}
\end{figure}

Figure \ref{fig:tb} presents maps of the peak brightness temperatures of
the observed molecular spectra (moment -2 maps in Miriad or moment 8 maps in CASA) in L1551 IRS 5.
The peak values in the C$^{18}$O, $^{13}$CO, CS, SO, and OCS maps
are 78, 92, 88, 106, and 86 K, respectively, located to the northwest of Source N.
% where the molecular emission traces the redshifted gas component of
% the rotating CBD (see the next subsection).
In the case of the thermal emission the observed brightness temperatures
should be the lower limit of the gas kinetic temperature,
and these results indicate that the gas kinetic temperature in the CBD exceeds $\gtrsim$ 100 K.
Overall, the western side of Source N in the CBD is brighter
than the eastern part. As the blueshifted molecular outflow is located to the
west, the western side corresponds to the surface of the CBD irradiated directly
from the radiation from Source N. The brightness temperature of the 0.9-mm continuum
emission in the CSD
associated with Source N exceeds $\sim$260 K, higher than that in Source S
($\sim$160 K). The bright molecular-line emission to the west of Source N
likely reflects the warm surface of the CBD irradiated from the intense radiation
from Source N.

Hereafter in this paper, the C$^{18}$O (3--2) and OCS (28--27) emission will be
adopted to investigate the gas motion in the CBD, as these molecular emission
appear to be the least affected by the associated molecular outflows.

% As described below, the newly-found velocity features
% in C$^{18}$O are also seen in the other lines at
% velocities sufficiently far away from the systemic velocity.
% Closer to the systemic velocity, the spatial-kinematic structure in the other lines is complicated by the effects
% of missing flux as well as an outflow (see Appendix A).

\subsection{Velocity Structures Traced by the C$^{18}$O and OCS Emission} \label{subsec:molvel}

In this subsection velocity structures of L1551 IRS 5 as traced by the
C$^{18}$O and OCS emission are presented.
From the symmetric center of the velocity features
identified in our Cycle 4 observations (see below), the systemic velocity of $v_{\rm sys}$ = 6.4 km s$^{-1}$ is adopted,
which is comparable to previous estimates \cite{fri02,tak11}.
% In our earlier ALMA study of L1551 IRS 5,
% Keplerian disk model fitting to the CBD derives the systemic velocity of
% $v_{\rm sys}$ = 6.4 km s$^{-1}$ \cite{yam20}. This velocity is consistent with
% the symmetric center of the velocity features identified
% in our Cycle 4 observations, and hereafter $v_{\rm sys}$ = 6.4 km s$^{-1}$ is adopted.
% From the investigation of the
% velocity channel maps and Position - Velocity (P-V) diagrams of the C$^{18}$O and OCS emission,
% $V_{LSR}$ = 6.4 km s$^{-1}$ is found to be the symmetric center of the observed velocity features.
% consistent with our earlier studies \cite{yam20}.
% Hereafter this velocity is adopted as the systemic velocity of L1551 IRS 5;
% $i.e.,$ $v_{\rm sys}$ = 6.4 km s$^{-1}$.

\subsubsection{Velocity Channel Maps} \label{subsubsec:velch}

\begin{figure}[ht!]
\figurenum{6}
\epsscale{1.2}
% \plotone{c18och3z7.pdf}
\plotone{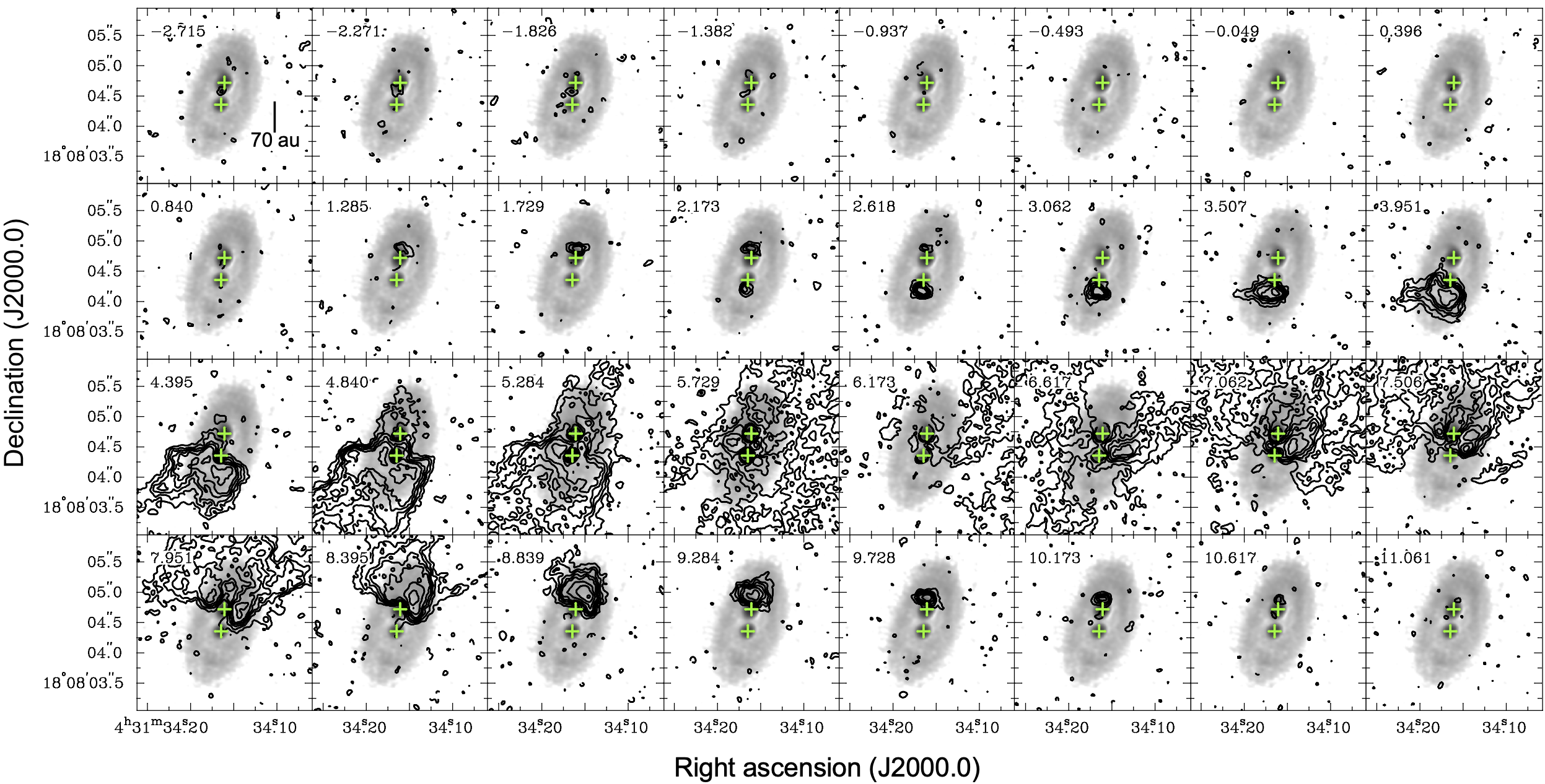}
\caption{Velocity channel maps of the C$^{18}$O (3--2) emission at a velocity
interval of 0.444 km s$^{-1}$ (contours),
superimposed on the 0.9-mm continuum image (gray; same as Figure \ref{fig:cont}a),
in L1551 IRS 5.
Contour levels are in steps of 3$\sigma$ until 15$\sigma$,
and then in steps of 5$\sigma$ (1$\sigma$ = 2.7 mJy beam$^{-1}$ = 2.3 K).
Numbers in the upper left corners denote the LSR velocity.
Crosses show the positions of the binary protostars.
\label{fig:c18och}}
\end{figure}

Figure \ref{fig:c18och} shows the velocity channel maps of the C$^{18}$O (3--2) line
in contours, superimposed on the 0.9-mm dust-continuum image of L1551 IRS 5 in gray scale.
% In the highest blueshifted velocities (1.176 -- 1.842 km s$^{-1}$), an emission component
In the highest blueshifted velocities (-2.7 km s$^{-1}$ -- -1.4 km s$^{-1}$),
an emission component to the south of Source N is seen. This component disappears
at $V_{LSR}$ = -0.9 km s$^{-1}$, and from $V_{LSR}$ = 1.3 km s$^{-1}$ to 2.6 km s$^{-1}$
a gas component to the north of Source N becomes evident. Following
this second blueshifted component
a gas component to the south of Source S appears from $V_{LSR}$ = 2.2 km s$^{-1}$.
This southern component extends to the east and
penetrates to the east of the protostellar binary in $V_{LSR}$ = 3.5 -- 5.3 km s$^{-1}$.
Around the systemic velocity (5.7 -- 7.1 km s$^{-1}$), the emission distribution
appears to be affected by the interferometric filtering effect, but
a butterfly shape of the emission can be discerned.
At the redshifted velocities of 7.1 -- 8.8 km s$^{-1}$
the C$^{18}$O emission is located to the north, and the emission
penetration to the west of the protobinary is also evident,
in contrast to the eastern penetration at the blueshifted side.
In the high redshifted velocities (9.3 -- 10.6 km s$^{-1}$)
% (9.175 -- 10.508 km s$^{-1}$)
the C$^{18}$O emission is seen to the north of Source N.
The location of this high-velocity redshifted emission is very close to that
of the high-velocity blueshifted (1.3 -- 2.6 km s$^{-1}$) emission.
No clear C$^{18}$O emission is seen beyond $V_{LSR}$ $\gtrsim$ 11 km s$^{-1}$.

Previous lower-resolution CS (7--6) observations
of L1551 IRS 5 have revealed the rotation motion of the CBD, where the northern part
is redshifted and the southern part blueshifted \cite{tak04,cho14}.
The southern blueshifted (2.2 -- 5.3 km s$^{-1}$) and northern redshifted (7.1 -- 10.6 km s$^{-1}$)
components as seen in the C$^{18}$O (3--2) emission likely trace the
same rotating gas motion in the CBD. Our higher-resolution
and higher-sensitivity ALMA observations have also unveiled the presence
of blueshifted eastern and redshifted western emission penetrations, as well as
the two highest blueshifted emission components to the south and north of Source N.

\begin{figure}[ht!]
\figurenum{7}
\epsscale{1.2}
% \plotone{ocsch.pdf}
\plotone{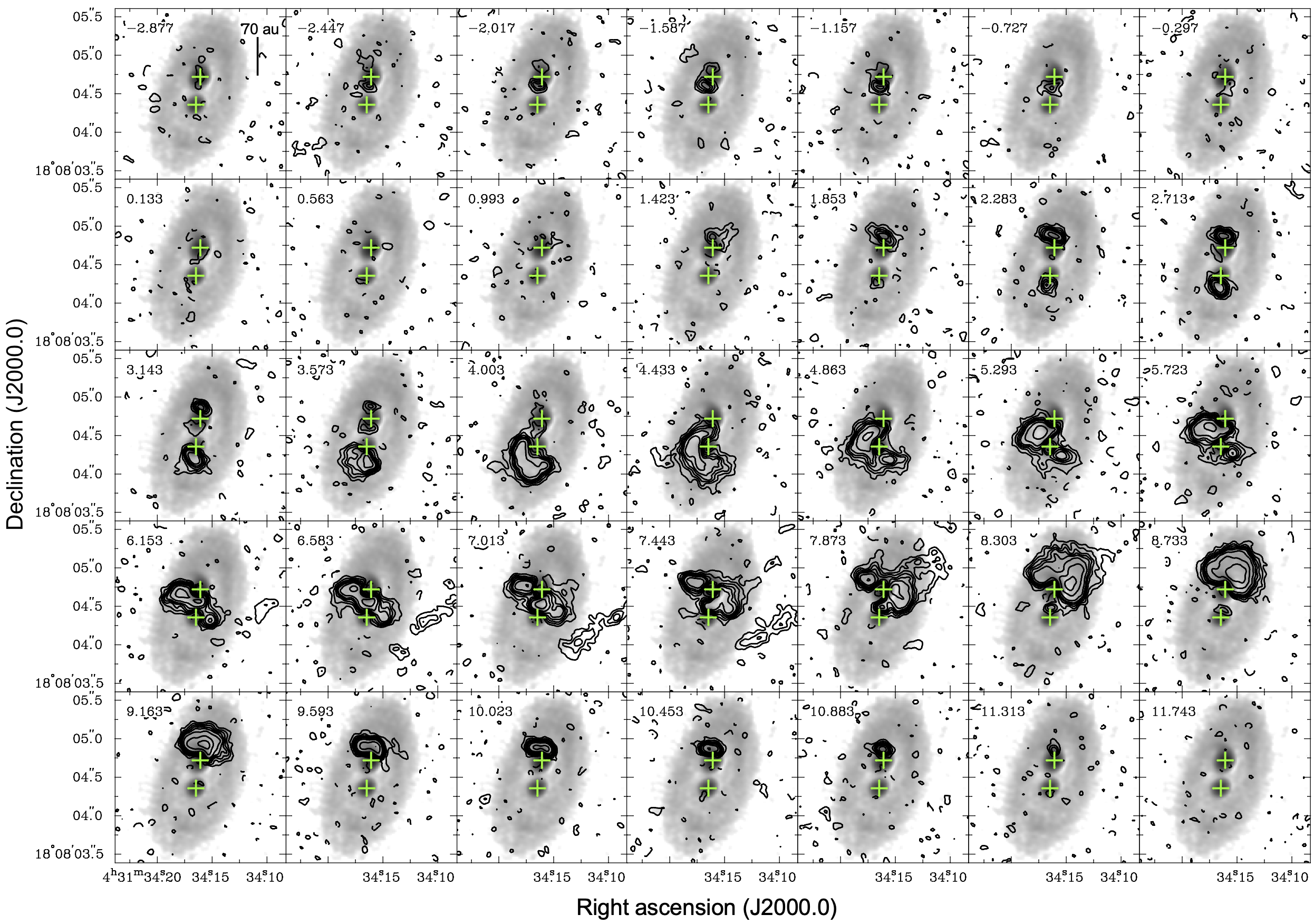}
\caption{Velocity channel maps of the OCS (28--27) emission at a velocity
interval of 0.43 km s$^{-1}$ (contours),
superimposed on the 0.9-mm continuum image (gray; same as Figure \ref{fig:cont}a),
in L1551 IRS 5. Contour levels are
3$\sigma$, 6$\sigma$, 9$\sigma$, 12$\sigma$, 15$\sigma$,
20$\sigma$, 30$\sigma$, 40$\sigma$, and then in steps of 20$\sigma$
(1$\sigma$ = 1.8 mJy beam$^{-1}$ = 1.6 K).
\label{fig:ocsch}}
\end{figure}

Figure \ref{fig:ocsch} shows the velocity channel maps of the OCS (28--27) emission.
As in the case of the C$^{18}$O emission, the highest blueshifted emission to the
south of Source N appears at $V_{LSR}$ = -2.4 -- -0.3 km s$^{-1}$.
At $V_{LSR}$ = -2.4 -- -1.2 km s$^{-1}$ a weak OCS emission to the north of Source N
is also seen. After these emission components disappear, from $V_{LSR}$ = 1.0 km s$^{-1}$
the OCS emission to the north of Source N appears, and then from
$V_{LSR}$ = 1.9 km s$^{-1}$ another OCS emission to the south of Source S emerges.
While the emission component to the north of Source N diminishes at
$V_{LSR}$ = 4.0 km s$^{-1}$, the emission component
to the south of Source S gradually shifts to the eastern side of the protobinary
from $V_{LSR}$ = 3.6 km s$^{-1}$ to 6.2 km s$^{-1}$.
Another emission component located to the south of Source N emerges
from 3.1 km s$^{-1}$. The peak location of this component
is systematically shifted to the south for the redder velocities, and
at $V_{LSR}$ = 8.7 km s$^{-1}$ the peak of this component is located
just to the north of Source S.
The C$^{18}$O counterpart of this component
can also be discerned in the C$^{18}$O velocity channel maps (Figure \ref{fig:c18och}).
An emission component to the southwest of Source S
also appears from $V_{LSR}$ = 4.9 km s$^{-1}$, and
in the redshifted velocities of
6.6 -- 8.3 km s$^{-1}$, this component is located to the west of the protostellar
binary.
In the highest redshifted velocities
(8.7 -- 11.3 km s$^{-1}$) the OCS emission is located to the north of Source N.
% and the peak location is closer to the location of Source N at the higher
% velocities.
The overall velocity features as seen in the OCS emission are thus
consistent with those in the C$^{18}$O emission, but in the OCS the emission component
located between Source N and Source S with a north (blueshifted)
to south (redshifted) velocity gradient is identified more clearly.

\subsubsection{P-V Diagrams} \label{subsubsec:pvs}

\begin{figure}[ht!]
\figurenum{8}
\epsscale{1.0}
% \plotone{c18opv160.pdf}
\plotone{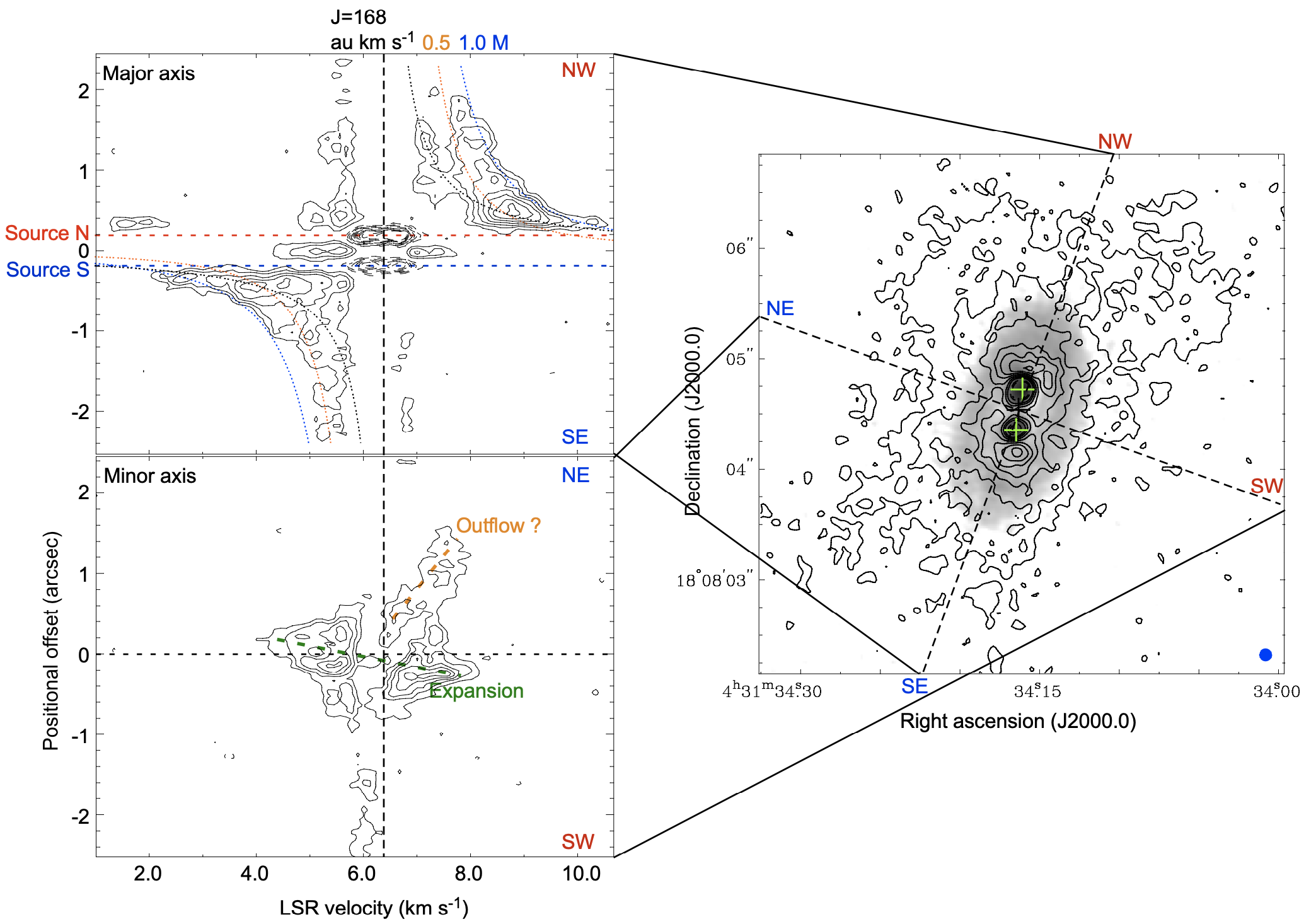}
\caption{Position-Velocity (P-V) diagrams of the C$^{18}$O (3--2) emission
along the major (upper-left panel; P.A. = 160$\degr$) and minor axes
(lower-left; P.A. = 70$\degr$) of
the CBD. The right panel shows the moment 0 map of the C$^{18}$O emission
(contour) superimposed on the 0.9-mm dust continuum image (gray),
as a guide map of the P-V cuts (dashed lines).
The origin of the positional axis is set to be the middle position
of Source N and Source S. Contour levels in the P-V diagrams are from 3$\sigma$
in steps of 2$\sigma$ (1$\sigma$ = 5.3 mJy beam$^{-1}$ = 4.5 K), and those in the moment 0 map
are in steps of 5$\sigma$ (1$\sigma$ = 5.5 mJy beam$^{-1}$ km s$^{-1}$ = 4.7 K km s$^{-1}$).
In the P-V diagram along the major axis,
upper and lower horizontal dashed lines denote the positions of Source N and Source S,
respectively.
Orange and blue curves represent the Keplerian rotation curves of the CBD
with the total binary mass of 0.5 $M_{\odot}$ and 1.0 $M_{\odot}$,
and black curves the rotational curve of the conserved specific angular
momentum ($\equiv j$) with $j$ = 168 au km s$^{-1}$,
where the adopted disk inclination angle $i$ is 60$\degr$.
In the P-V diagram along the minor axis, a horizontal dashed line denotes
the origin of the P-V diagram, and green and orange dashed lines delineate
the detected velocity features. A vertical dashed line represents the
systemic velocity of 6.4 km s$^{-1}$.
\label{fig:c18opv}}
\end{figure}

Figure \ref{fig:c18opv} shows the Position-Velocity (P-V) diagrams of the C$^{18}$O emission
along the major (P.A. = 160$\degr$) and minor axes (P.A. = 70$\degr$) of the CBD
in L1551 IRS 5 \footnote{For a clear presentation the velocity range in
these P-V diagrams is adjusted and the highest blueshifted component to the south
of Source N is not included. The P-V diagrams including the highest blueshifted
component are shown in Figure \ref{fig:nspv}.}.
In these P-V diagrams the zero position is set to be the middle position
between Source N and Source S, which is considered to be the dynamical center of the CBD
on the assumption of equal protostellar masses.
Along the major axis, the primary emission components are
northwestern redshifted and southeastern blueshifted components.
% which reflect the rotating motion of the CBD.
The northwestern redshifted component exhibits
higher velocities at the position closer to Source N, and the southeastern blueshifted
component closer to Source S, suggesting the presence of spin-up rotation.
The middle position between Source N and Source S (zero position
in the P-V diagram) appears to be consistent with the dynamical center of rotation.
% implying the equal binary mass.

Previous interferometric observations of L1551 IRS 5 have claimed that
the CBD shows Keplerian rotation with the central stellar mass (total binary mass)
of $\sim$0.5 $M_{\odot}$ and the outermost radius of $\sim$70 au \cite{tak04,cho14},
and that outside the Keplerian CBD
there is a $\sim$2500 au scale, infalling and rotating protostellar envelope with
the rotational specific angular momentum of $\sim$168 au km s$^{-1}$ \cite{mom98}.
To discuss the velocity structures observed with the new ALMA observations,
the Keplerian rotation curves with the central stellar masses of 0.5 and 1.0 $M_{\odot}$
and the rotation curve with the conserved specific angular momentum of 168 au km s$^{-1}$
are drawn on the P-V diagram along the major axis. At a radius of $\sim$70 au
(=$\pm$0$\farcs$5), the emission ridges appear to be
consistent with the Keplerian rotation curve from a central stellar mass of
0.5 $M_{\odot}$. As the positions are closer to the locations of Source N or Source S,
the velocity increase of the
observed emission is even more than that of the Keplerian rotation curve of 0.5 $M_{\odot}$.
% The presence of the highest redshifted and blueshifted velocities
% in the close vicinity of Source N and Source S, respectively, likely indicate
% the transition from the global Keplerian rotation in the CBD to rotations
% in the individual CSDs.
%
% --> Simulation でも見えている、いない？
% The high-velocity blueshifted component to the north of Source N,
% as seen in the velocity channel maps (Figure \ref{fig:c18och}),
The gas component between Source N and Source S as identified in the OCS velocity channel maps
(Figure \ref{fig:ocsch}) is also seen in the C$^{18}$O P-V diagram along the major axis.

In the P-V diagram along the minor axis, there are northeastern blueshifted and southwestern redshifted
components. The positions of these two components are within the spiral arms
as seen in the 0.9-mm dust-continuum emission (see Figure \ref{fig:c18opv} right).
As the near and far sides of the CBD are to the northeast and southwest, respectively,
the northeastern blueshifted and southwestern redshifted velocities correspond to
the blueshifted velocity on the near side and the redshifted velocity on the far side.
Such a velocity feature can be interpreted as the radial expanding gas motion in the CBD
(green dashed line in Figure \ref{fig:c18opv}). There is another emission component
to the northeast, which exhibits higher redshifted velocities further from the center
(orange dashed line in Figure \ref{fig:c18opv}).
This emission component corresponds to an extended gas component located to the northeast outside
the CBD at $V_{LSR}$ = 6.6 -- 8.0 km s$^{-1}$ (see Figure \ref{fig:c18och}).
One interpretation of this component is an outflow contamination, perpendicular to the CBD plane.

\begin{figure}[ht!]
\figurenum{9}
\epsscale{1.0}
% \plotone{ocspv160.pdf}
\plotone{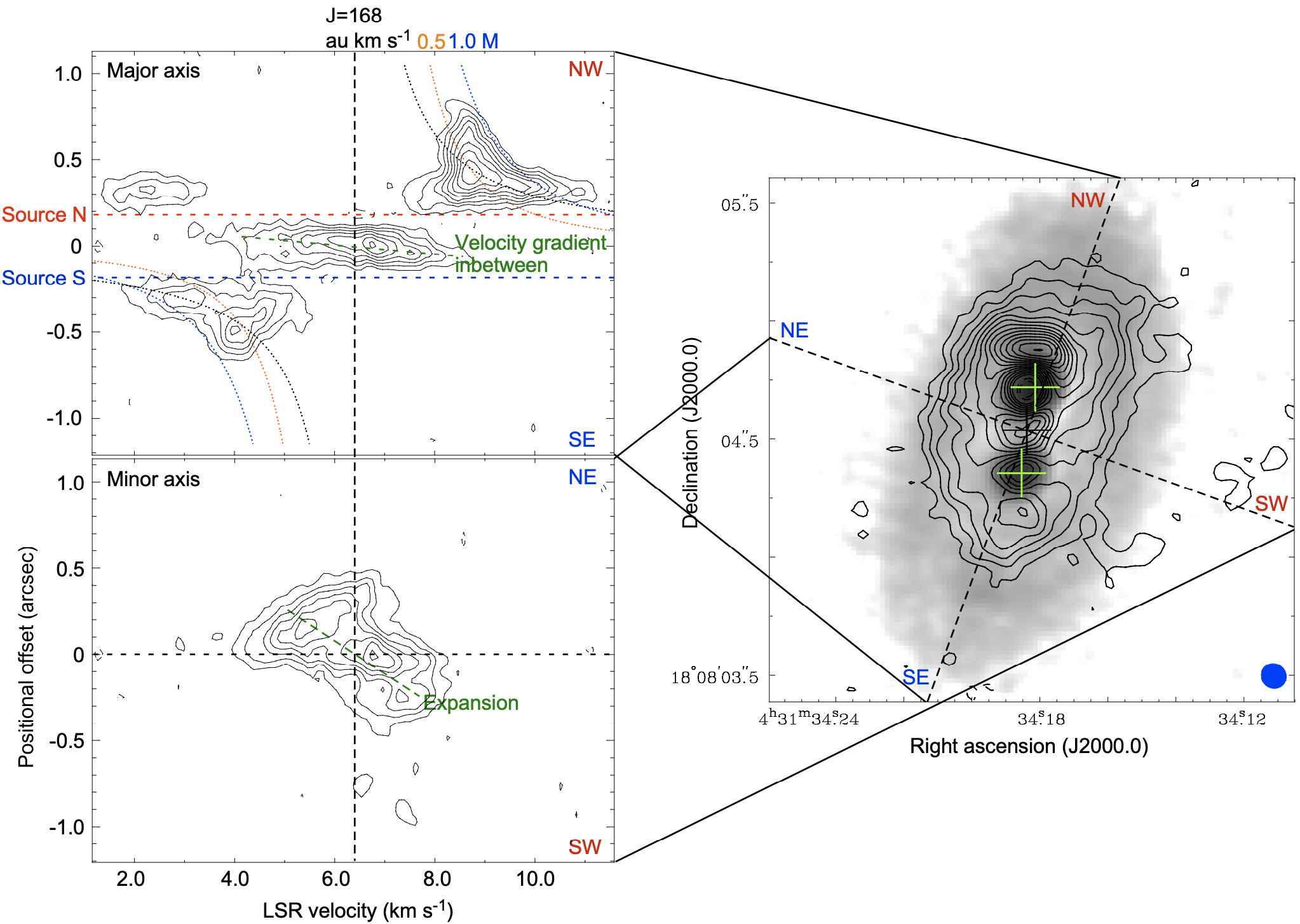}
\caption{P-V diagrams of the OCS (28--27) emission
along the major (upper-left~panel) and minor axes (lower-left) of
the CBD. The right panel shows the moment 0 map of the OCS emission
(contour) superimposed on the 0.9-mm dust continuum image (gray),
as a guide map of the P-V cuts (dashed lines).
The origin of the positional axis is set to be the middle position
of Source N and Source S. Contour levels in the P-V diagrams are
in steps of 3$\sigma$ (1$\sigma$ = 3.5 mJy beam$^{-1}$ = 3.1 K),
and those in the moment 0 map are in steps of 5$\sigma$
(1$\sigma$ = 3.7 mJy beam$^{-1}$ km s$^{-1}$ = 3.3 K km s$^{-1}$).
In the P-V diagram along the major axis,
upper and lower horizontal dashed lines denote the
positions of Source N and Source S, respectively.
Orange and blue curves represent the Keplerian rotation curves of the CBD
with the total binary mass of 0.5 $M_{\odot}$ and 1.0 $M_{\odot}$,
and black curves the rotational curve of the conserved specific angular
momentum ($\equiv j$) with $j$ = 168 au km s$^{-1}$.
In the P-V diagram along the minor axis, a horizontal dashed line denotes
the origin of the P-V diagram. Green dashed lines delineate
the detected velocity features. A vertical dashed line represents the
systemic velocity of 6.4 km s$^{-1}$.
\label{fig:ocspv}}
\end{figure}

Figure \ref{fig:ocspv} shows the same as that of Figure \ref{fig:c18opv} but for
the OCS (28--27) emission. As the extent of the OCS emission is much
smaller than that of the C$^{18}$O emission and even smaller than
that of the 0.9-mm dust-continuum emission, the OCS emission should
selectively trace the gas motions in the innermost part of the CBD.
In the P-V diagram along the major axis,
the northwestern redshifted and southeastern blueshifted components with
the spin-up rotating signatures are identified.
% The symmetric axes of these components are consistent with
% the the middle position between Sources N and S (zero position
% in the P-V diagram) along the positional axis and $v_{\rm sys}$ = 6.4 km s$^{-1}$
% along the velocity axis.
As in the case of the C$^{18}$O emission, the velocity increase of the
observed emission is even more than that of the Keplerian rotation
curve of 0.5 $M_{\odot}$ closer to the locations of the protobinary.
The high-velocity blueshifted component to the north of Source N
is also clearly identified in the P-V diagram. A gas component located
between Source N and Source S with a blueshifted (north) to redshifted (south)
velocity gradient, which is seen in the velocity channel maps
(Figure \ref{fig:ocsch}), is also identified. In the P-V diagram along
the minor axis, the blueshifted emission is located predominantly to the
northeast and the redshifted emission to the southwest, consistent
with the radial expanding motion in the CBD.

\section{Discussion} \label{sec:dis}
\subsection{Nature of the Observed Structures and Gas Motions in L1551 IRS 5} \label{subsec:cbdirs5}

Our ALMA Cycle 4 observations of L1551 IRS 5 have unveiled a two spiral-arm structure
of the CBD in the 0.9-mm dust-continuum emission, and rotating and expanding gas motions
of the CBD in the C$^{18}$O (3--2) and OCS (28--27) emission.
There are also other gas components, $i.e.$,
the gas component with the north (blueshifted) - south (redshifted) velocity gradient
located between Source N and Source S and
the high-velocity blueshifted emission components to the north and south of Source N.
To discuss the nature of these observed features, we performed hydrodynamical numerical simulations
using the adaptive mesh refinement (AMR) code SFUMATO \cite{mat07}.
Details of our numerical simulations are described in \citealp[][]{mat19}.
Our simulations have been adopted to interpret the structure and gas motion of the
CBD in L1551 NE, and have successfully reproduced the observed two-arm spirals,
$m$ = 1 mode, and the expanding gas motion \cite{tak14,tak17}.
To discuss the observational results of L1551 IRS 5, physical parameters of the simulations
are tailored to match with those of L1551 IRS 5 (Table \ref{theory}).
In particular, the gas kinetic temperature in our isothermal simulation is set to be
$T_K$ = 100 K, to account for the inferred high temperature (see Figure \ref{fig:tb}).
We chose the simulation result at the 58.7 orbital period of the binary ($\sim$3.3 $\times$ 10$^4$ yr)
after the start of the simulation, where one orbital period is $\sim$565 yr.
The simulation result at the 58.7 orbital period appears to resemble the observed 0.9 mm dust-continuum image.
The output from the numerical simulation was transferred to the radiative transfer
calculation, and the theoretically predicted 0.9-mm dust-continuum and C$^{18}$O (3--2)
images were constructed on the assumption of the LTE condition.
Details of the radiative transfer calculation are described by \citealp[][]{tak14}.
% \citeyearpar{tak14}.
We measured the radial profile of the sum of the observed 0.9-mm
brightness temperature and the peak brightness temperature of the C$^{18}$O
spectra ($i.e.,$ sum of Figures \ref{fig:cont}a and \ref{fig:tb}),
and found that the temperature profile can be approximated as
$T(r) = \max\left[ 62~{\rm K} \left(\frac{r_N}{100~{\rm au}}\right)^{-0.5},
48~{\rm K} \left(\frac{r_S}{100~{\rm au}}\right)^{-0.5} \right]$,
where $r_N$ and $r_S$ indicate the distance from Source N and Source S, respectively.
This temperature profile was adopted in the radiative transfer calculation.
Note that in the numerical simulation the gas kinetic temperature is assumed to be uniform,
but at the stage of the radiative transfer the temperature profile is incorporated.
% The opacity of the 0.9-mm continuum emission is set to be uniform \cite{tak14}.
% and the conversion factor
% from the numerical results to the 0.9-mm continuum flux densities is chosen to approximately
% match the theoretical and observational 0.9-mm flux densities.
The C$^{18}$O abundance was adjusted to approximately match the observed and model
C$^{18}$O (3--2) intensities,
and the adopted value is $X$(C$^{18}$O) = 5.1 $\times$ 10$^{-7}$,
which is a factor of 3 higher than the canonical value of 1.7 $\times$ 10$^{-7}$ \cite{cra04}.
Parameters of these modelings are summarized in Table \ref{theory}.
Then CASA observing simulations were performed to create the simulated visibility data
for the model images with the same antenna configuration,
hour angle coverage, bandwidth and frequency resolution,
and integration time as those of the real observations.
In the CASA simulations the weather parameters
were also adjusted to approximately give similar noise levels to those of the real data.
The simulated theoretical images with the same imaging methods
as those of the real data were then made.

%%%%%%%%%%%
% Table 3 %
%%%%%%%%%%%
\clearpage
\begin{deluxetable}{ll}
\tabletypesize{\scriptsize}
\tablecaption{Parameters for the Theoretical Model of L1551 IRS 5 \label{theory}}
\tablewidth{0pt}
%\tablewidth{250pt} % use this in emulateapj
\tablehead{\colhead{Parameter} &\colhead{Value}}
\startdata
Computer & ATERUI II in NAOJ CfCA\\
Simulation Code & 3D AMR Code \cite[SFUMARTO;][]{mat07,mat19} \\
% Execution Time & \textcolor{red}{35 hours (512 Cores)} \\
Simulation Box & (1296 au)$^2$ $\times$ 648 au\\
Highest Resolution &0.32 au \\
Radius of Sink Particles &1.3 au\\
% Boundary Radius &\textcolor{red}{1740 au}\\
Image Pixel Size & 2 au \\
Binary Separation         & 54 au \\
Centrifugal Radius of the Injected Gas$\tablenotemark{a}$ & 64 au\\
Disk Position Angle$\tablenotemark{a}$ & 160$^{\circ}$\\
Disk Inclination Angle$\tablenotemark{a}$      & 60$^{\circ}$  \\
Systemic Velocity$\tablenotemark{a}$  & 6.4 km s$^{-1}$ \\
Total Binary Mass$\tablenotemark{a}$                 &  0.5 $M_{\odot}$    \\
Binary Mass Ratio$\tablenotemark{b}$ & 1.0  \\
% Gas Number Density at the Boundary & \textcolor{red}{1.5 $\times$ 10$^{5}$ cm$^{-3}$}\\
% Mean Molecular Weight & 2.3 \\
Temperature Profile$\tablenotemark{c}$ &
% \revisetm{
$T(r) = \max\left[ 62~{\rm K} \left(\frac{r_N}{100~{\rm au}}\right)^{-0.5},
48~{\rm K} \left(\frac{r_S}{100~{\rm au}}\right)^{-0.5} \right]$
% }
\\
% Dust Opacity$\tablenotemark{c}$   &
% $\kappa_{0.9~mm} = 0.053~\mathrm{cm}^{2}~\mathrm{g}^{-1}$ \\
C$^{18}$O Abundance & $5.1 \times 10^{-7}$  \\
\enddata
\tablenotetext{a}{\citealp[][]{cho14}; This work.}
\tablenotetext{b}{\citealp[][]{lim06,lim16a}.}
\tablenotetext{c}{Radii $r_N$ and $r_S$ indicate the distance from Source N and that from
Source S, respectively.}
% \tablenotetext{c}{Crapsi et al. (2004).}
\end{deluxetable}

\subsubsection{Structures and Gas Motions in the Spiral Arms}
\begin{figure}[ht!]
\figurenum{10}
\epsscale{1.2}
% \plotone{contobsmodv7.pdf}
\plotone{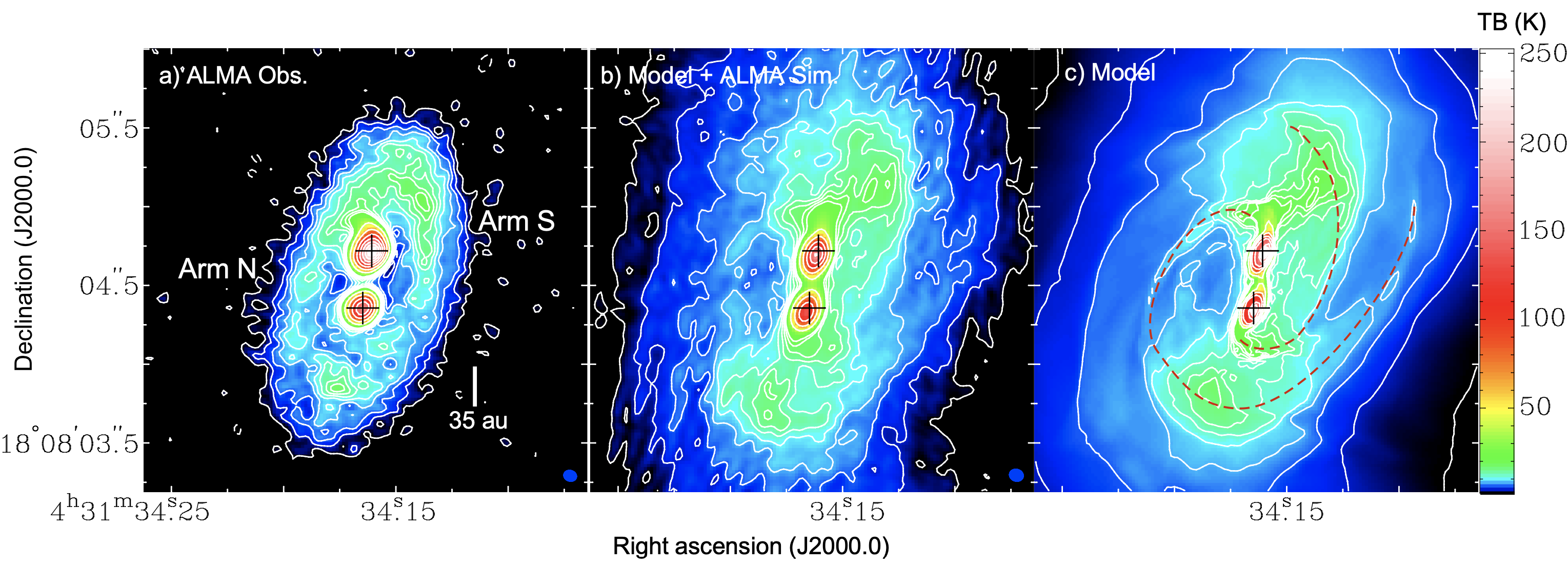}
\caption{Comparison of the observed (panel a) and model 0.9-mm dust-continuum
images of L1551 IRS 5 (b,c).
% a) 0.9-mm dust-continuum image of L1551 IRS 5 observed with ALMA.
% Upper and lower crosses indicate the centroid positions of
% the 2-dimensional Gaussian fittings to the central two dusty components,
% which we regard as the positions of Sources N and S.
% A filled ellipse at the bottom-right corner shows the synthesized beam
% (0$\farcs$0788$\times$0$\farcs$0652; P.A. = 71.7$\degr$).
% b), c) Theoretically-predicted 0.9-mm dust-continuum images of L1551 IRS 5.
We performed
the radiative transfer calculation with the gas distribution computed from our 3-D hydrodynamic
model to produce the theoretical image shown in panel c). Then we conducted the ALMA observing
simulation to make the theoretically-predicted ALMA image shown in panel b).
Contour levels are common in all the panels, and in steps of 3$\sigma$ until 30$\sigma$,
then 60$\sigma$, 100$\sigma$, and then in steps 50$\sigma$ (1$\sigma$ = 0.67 K).
The color scale denotes the brightness temperature. In panel c the present two spiral arms
are delineated by red dashed curves.
\label{fig:cont3}}
\end{figure}

Figures \ref{fig:cont3} (a), (b), (c) show the observed 0.9-mm continuum image,
and the theoretical continuum images after and before the CASA observing simulation,
respectively. Two spiral arms can be discerned in the theoretical images
(dashed curves in Figure \ref{fig:cont3}c).
One arm starts from the north of Source N, points toward the east,
and then curls through the south to the west. The other arm starts from the south
of Source S, points to the west, and then curls toward the northwest.
Such features in the model appear to be consistent with the observed spiral arms;
$i.e.$, the arm starting from the north of Source N corresponds to Arm N
and the arm from the south of Source S to Arm S.
On the other hand, the extent of the model CBD as a whole is larger than that of the observed image,
and the observed CBD as seen in the 0.9-mm dust-continuum emission is rather confined.
The extension of the CBD in the numerical simulations is due to the high temperature
of the gas and angular momentum transport by the gravitational torque of the binary stars.
A possible reason for the observed confinement of the CBD is the magnetic field,
which is not included in the numerical simulation.

\begin{figure}[ht!]
\figurenum{11}
\epsscale{1.2}
% \plotone{c18omodx3ch3z7.pdf}
% \plotone{c18omodx3ch3z7.png}
\plotone{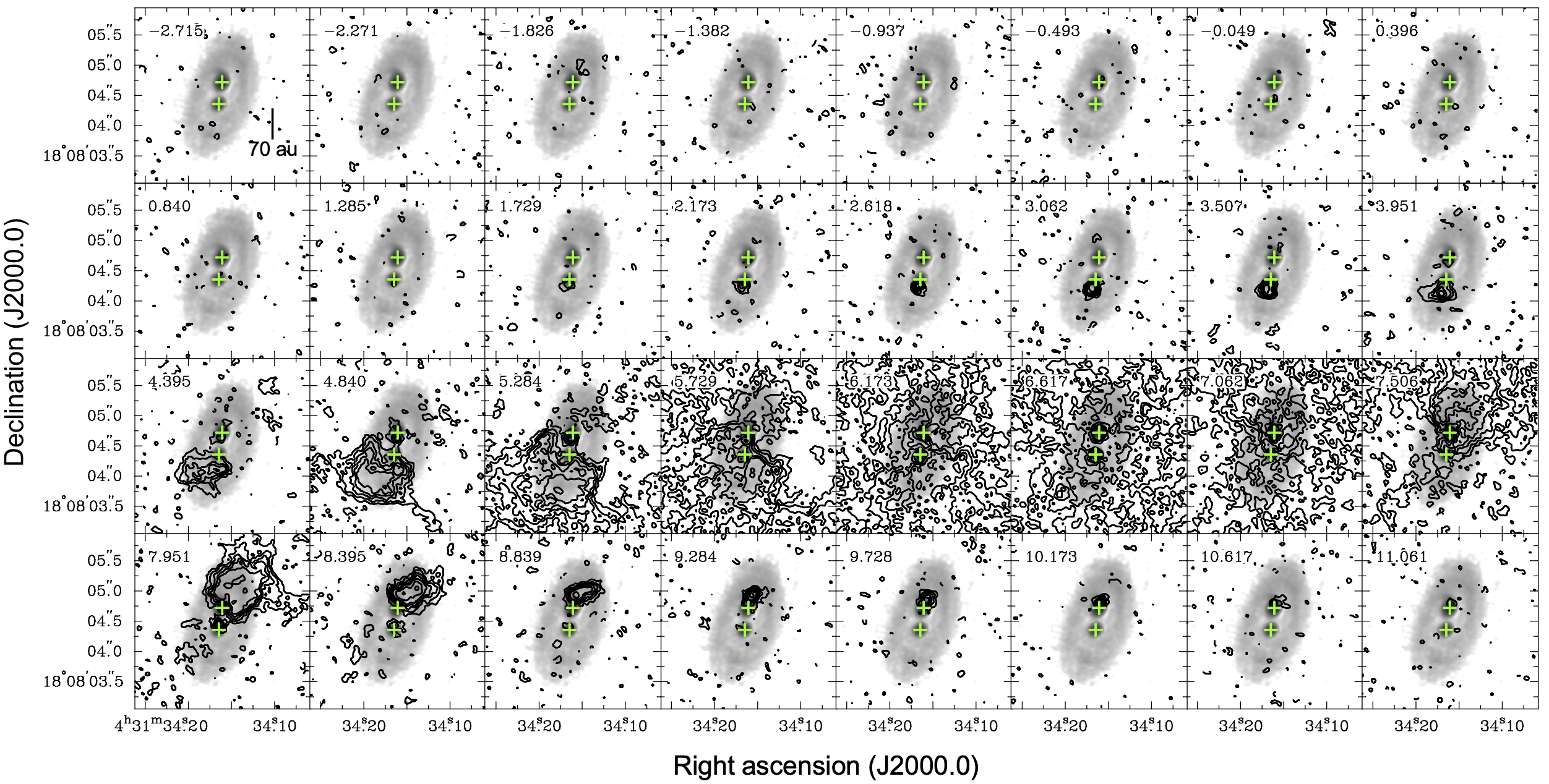}
\caption{Theoretically-predicted velocity channel maps of the C$^{18}$O ($J$=3-2) emission
(contours) superimposed on the observed 0.9-mm dust-continuum emission (gray scale) in L1551 IRS 5,
calculated from our hydrodynamic and radiative transfer calculations and the ALMA observing simulation.
Contour levels and symbols are the same as those in Figure \ref{fig:c18och}.
\label{fig:modch}}
\end{figure}

Figure \ref{fig:modch} shows the simulated theoretical velocity channel maps
of the C$^{18}$O (3-2) line (contours)
superimposed on the observed 0.9-mm dust-continuum image (grey scale),
the model counterpart of Figure \ref{fig:c18och}.
In contrast to the observed C$^{18}$O and OCS velocity channel maps,
the high-velocity blueshifted components to the south of Source N
(-2.7 km s$^{-1}$ -- -1.4 km s$^{-1}$) and to the north of Source N
(1.3 km s$^{-1}$ -- 2.6 km s$^{-1}$) are not seen in the model velocity
channel maps. This result implies that the origin of these observed
highly blueshifted emission components should be distinct from the CBD model.
On the other hand, the model counterpart of the observed blueshifted emission
to the south of Source S is seen from $V_{\rm LSR}$ = 1.7 km s$^{-1}$.
From $V_{\rm LSR}$ = 3.5 km s$^{-1}$ this component extends eastward,
and the C$^{18}$O emission penetrates to the east of the protobinary
until $V_{\rm LSR}$ = 5.3 km s$^{-1}$. Such a velocity feature is also consistent with
the observation as described in section 3.3.1.
On the redshifted side
a similar velocity feature is also seen both in the observed and model velocity channel maps.
At the highly redshifted velocities of 9.7 -- 10.6 km s$^{-1}$
compact C$^{18}$O emission located to the north of Source N is present,
and at $V_{\rm LSR}$ = 7.5 -- 9.3 km s$^{-1}$ this component extends westward
and penetrates to the west of the protobinary.
% Around the systemic velocities (5.7 -- 7.1 km s$^{-1}$) the effect of the missing flux
% prevents us from direct comparison between the observed and model velocity channel maps.

\begin{figure}[ht!]
\figurenum{12}
\epsscale{1.0}
\plotone{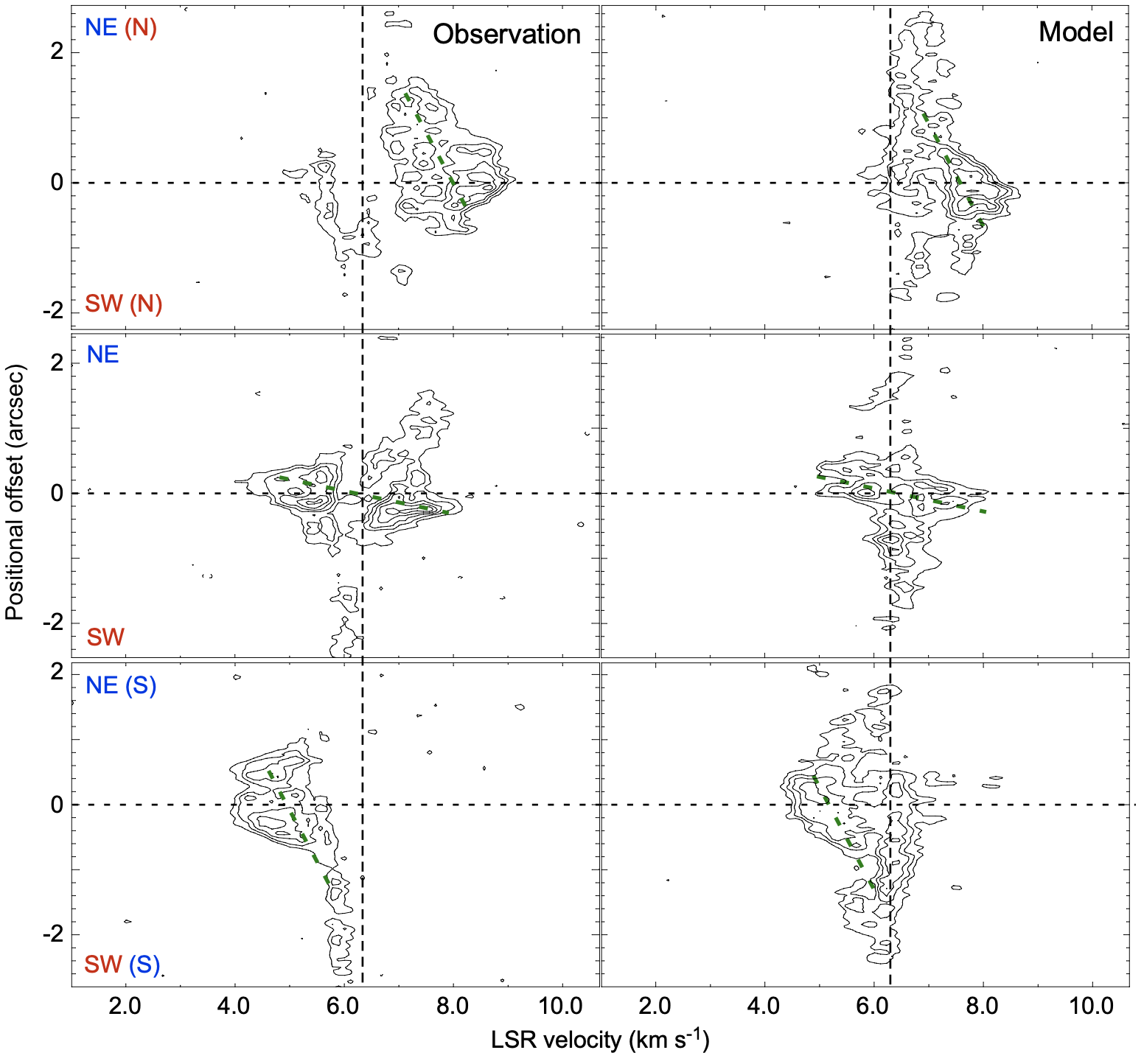}
\caption{Comparison of the observed (left panels) and
model (right) P-V diagrams of the C$^{18}$O (3--2) emission in L1551 IRS 5.
Top, middle, and bottom panels show the P-V diagrams
along the cuts parallel to the minor axis of the CBD (P.A. = 70$\degr$) passing through
+0$\farcs$75, 0$\arcsec$, and -0$\farcs$75 offsets
from the middle position of Source N and Source S, respectively.
Contour levels are from 3$\sigma$
in steps of 2$\sigma$ (1$\sigma$ = 5.3 mJy beam$^{-1}$ = 4.5 K).
Horizontal and vertical dashed lines denote the origin of the P-V diagrams
and the systemic velocity of 6.4 km s$^{-1}$, respectively.
Green dashed lines delineate the detected velocity features.
\label{fig:c18opvmin}}
\end{figure}

Figure \ref{fig:c18opvmin} compares the observed (left panels) and model (right)
P-V diagrams of the C$^{18}$O (3--2) emission
along the cuts parallel to the minor axis
of the CBD (P.A. = 70$\degr$) passing through +0$\farcs$75 (top panel),
0$\arcsec$ (midddle), and -0$\farcs$75 offsets (bottom) from
the middle position between the protobinary.
The P-V diagrams at the $\pm$0$\farcs$75 offsets are across
the spiral arms as seen in the 0.9-mm dust continuum emission,
while that at the 0$\arcsec$ offset is along the minor axis of the CBD
(same as that in Figure \ref{fig:c18opv}).
As already discussed the P-V diagram along the minor axis of the CBD shows
the northeast (blueshifted) to southwest (redshifted) velocity gradient
(green dashed line in Figure \ref{fig:c18opvmin}),
which can be interpreted as the expanding gas motion of the CBD.
The model P-V diagram along the minor axis reproduces such an expanding
gas motion.
If the gas motion in the CBD is azimuthal only, the pattern
of the line-of-sight (LOS) velocity should be symmetric with respect
to the disk major axis, and the P-V diagrams across the arms
parallel to the minor axis should show symmetric patterns
with respect to the major axis \cite{tak14}.
The P-V diagrams across the arms show, however, rather skewed emission
distributions, and northeast (blueshifted) to southwest (redshifted) velocity gradients
(green dashed lines), which are also reproduced with the model P-V diagrams.
These velocity features reflect the eastern blueshifted and western redshifted emission
penetrations seen in the observed (Figures \ref{fig:c18och}, \ref{fig:ocsch})
and the model velocity channel maps (Figure \ref{fig:modch}).
These results indicate that the spiral arms have expanding motions,
as well as rotation as seen in the P-V diagrams along the major axis
(Figures \ref{fig:c18opv} and \ref{fig:ocspv}).

Our numerical simulations, as well as previous theoretical models
\cite{bat00,gun02,och05,han10,shi12,dem15,you15,pri18}, predict a two-arm spiral pattern in CBDs.
The arms extend from the binaries through the L2 and L3 Lagrangian points
in the Roche potential.
Our simulations show that the spirals co-rotate with the binary system, and thus
the spiral pattern rotates faster than the material in the CBD.
Outside the L2 and L3 Lagrangian points,
the spirals in the CBD are the regions where the nonaxisymmetric gravitational torques from the binary
impart angular momenta and drive faster rotation than the local Keplerian velocity.
Material in the spirals thus expands outward.
After the spiral pattern passes through the location of the material, it loses angular momenta
and falls inward. The net effect is gas infall from the CBD to the CSDs.

The observed radial expansion in the CBD
of L1551 IRS 5 can thus be interpreted as a result of
the non-axisymmetric gravitational torques of the binary.
In the neighboring protostellar binary L1551 NE, a similar
expanding gas motion in the arms has also been identified % as well as the global rotation of the CBD
\cite{tak14,tak17}.
In the case of L1551 NE infalling gas components in the inter-arm regions of the CBD
are also identified \cite{tak14,tak17}, which are not clearly identified in the CBD of L1551 IRS 5.
The apparent absence of infall in the CBD of L1551 IRS 5 should not be regarded, however,
as the real absence of the infall in the CBD. Our numerical simulation indeed shows
presence of the infalling motion. To observationally identify infall,
sufficiently large area of the infalling region must reside
along the minor axis of the CBD.
In the case of L1551 IRS 5 the projected locations of the protobinary are closely
aligned along the major axis of the CBD. As shown in Figure 5 of \citealp[][]{mat19},
in such a configuration most
of the infalling regions resides along the major axis.
On the other hand, in L1551 NE the locations of the protobinary are well tilted
from the major axis of the CBD, and the large infalling area can pass through
the minor axis of the CBD. Furthermore,
in the case of L1551 NE the size of the CBD itself is larger and the minor axis
can pass through the sufficiently large inter-arm regions (see Figure \ref{fig:neirs5}).
% binary and the spiral configurations are tilted with respect to the
% CBD major axis, and thus the CBD minor axis passes through the sufficiently large inter-arm
These observational biases cause the apparent absence of infall in L1551 IRS 5
in spite of the presence of infall in L1551 NE.

\subsubsection{Gas Motion inside the Hill Radii}

\begin{figure}[ht!]
\figurenum{13}
\epsscale{1.2}
\plotone{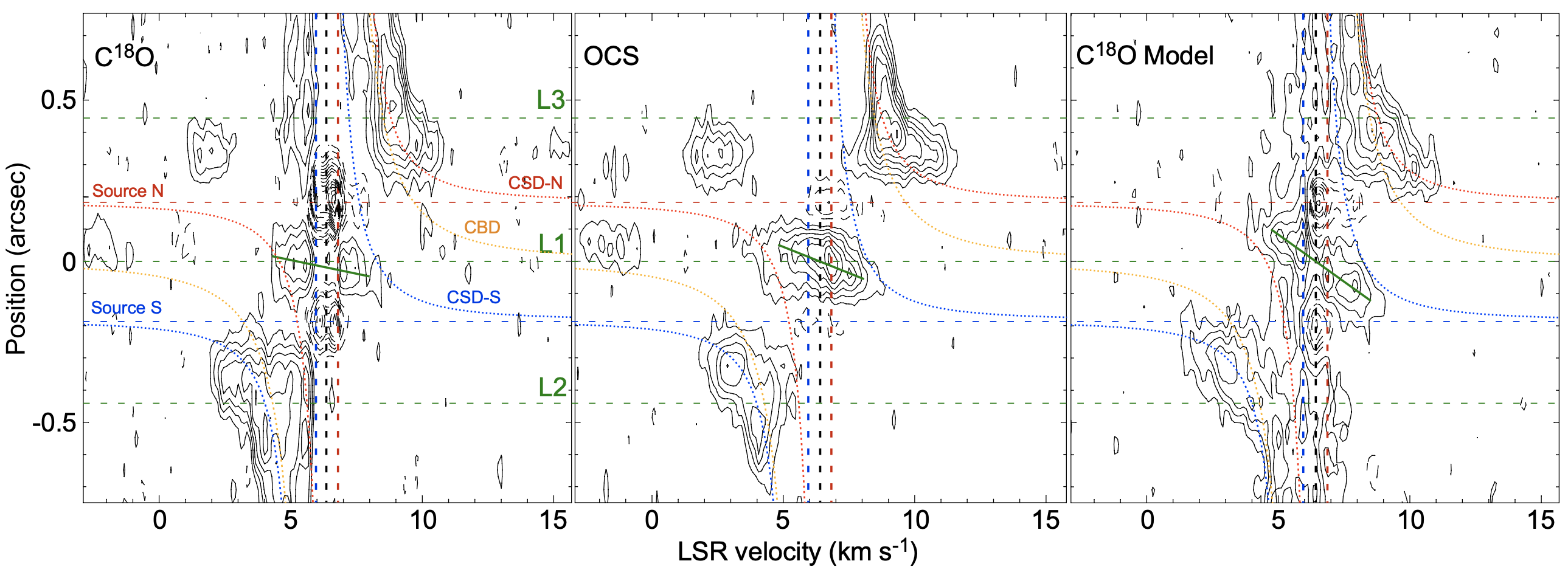}
\caption{Observed P-V diagrams of
C$^{18}$O (3--2) (left panel) and OCS (28--27) emission (middle),
and the model P-V diagram of C$^{18}$O (3--2) emission (right)
passing through Source N and Source S (P.A. = 171$\degr$).
Contour levels in the observed and model C$^{18}$O P-V diagrams are
in steps of 2$\sigma$ (1$\sigma$ = 5.3 mJy beam$^{-1}$ = 4.5 K), and those in the OCS P-V diagram
in steps of 3$\sigma$ (1$\sigma$ = 3.5 mJy beam$^{-1}$ = 3.1 K).
Red and blue horizontal dashed lines denote the positions of Source N and Source S,
and green horizontal dashed lines the Lagrangian points as labeled.
Red, black, and blue vertical dashed lines show
the anticipated LSR velocity of Source N (= 6.85 km s$^{-1}$), the systemic velocity (= 6.4 km s$^{-1}$),
and the anticipated LSR velocity of Source S (= 5.95 km s$^{-1}$), respectively.
Orange curves represent the Keplerian rotation curves of the CBD
with the total binary mass of 0.5 $M_{\odot}$, and the red and blue curves
the Keplerian rotation curves of the CSDs around Source N and Source S with
the individual stellar mass of 0.25 $M_{\odot}$, respectively.
Solid green lines denote
the detected velocity gradients around the L1 point.
\label{fig:nspv}}
\end{figure}

Figure \ref{fig:nspv} shows the observed P-V diagrams of the
C$^{18}$O (left panel) and OCS emission (middle), and the model C$^{18}$O P-V diagram (right)
in the close vicinity of Source N and Source S passing through them.
The position angle of the P-V cut (P.A. = 171$\degr$)
is close to that of the major axis of the CBD, and from these zoomed P-V diagrams the azimuthal
gas motions near the individual protostellar sources can be investigated.
On the assumption of a circular Keplerian orbit of the binary system
co-planar to the CBD with the inclination angle of $i$ = 60$\degr$ and position angle
of $\theta$ = 160$\degr$, and a total binary mass of 0.5 $M_{\odot}$ with mass ratio of unity
\cite{lim06,lim16a,cho14}, the line-of-sight velocities of Source N and Source S
are calculated to be $V_{LSR}$ = 6.85 km s$^{-1}$ and 5.95 km s$^{-1}$, respectively.
These velocities, as well as the positions of the protobinary, must be the centers
of the CSD rotations. On these positional and velocity centers
the anticipated Keplerian rotation curves of the CSDs around Source N (red curves) and Source S (blue curves)
are drawn in Figure \ref{fig:nspv}, as well as the Keplerian rotation curves of the CBD (orange).
As already described in Figures \ref{fig:c18opv} and \ref{fig:ocspv},
at the outer radii the northwestern redshifted and southeastern blueshifted components
originate from the rotating CBD.
In the closer vicinities of Source N and Source S these emission components extend
to even redder and bluer velocities ($\gtrsim v_{sys} \pm$3 km s$^{-1}$), respectively,
seen in both the observed and model P-V diagrams.
The degree of the velocity increase in the high redshifted and blueshifted
velocities seems to better match with the Keplerian rotation of the CSDs
around Source N and Source S, respectively, than that of the CBD.
This result implies the transition from the outer CBD to the inner CSDs.
In the close vicinity of the individual binary protostars
the gravitational field from the individual single star should be dominant
and control the rotational gas motions in the individual CSDs.
Such transitions should take place around the L2 and L3 Lagrangian points,
and the outermost radius of the gravitational field from the single star is called as
Hill radius. On the assumption of the equal masses of the individual stars,
the distance between the L1 and L2, L3 Lagrangian points should be $\sim$1.2$a$,
where $a$ denotes the binary separation.
The observed transitions from the CBD to the CSDs
indeed appear to occur around the inferred L2 and L3 Lagrangian points (green horizontal dashed lines in
Figure \ref{fig:nspv}).

In the observed velocity channel maps of the OCS emission (Figure \ref{fig:ocsch}),
there is a gas component with the north (blueshifted) - south (redshifted) velocity gradient
located between the protobinary. The observed C$^{18}$O counterpart can also be identified,
and these observed components are clearly seen in Figure \ref{fig:nspv}.
In the model velocity channel maps of Figure \ref{fig:modch}, a gas component to the south of Source N
emerges from $V_{\rm LSR}$ = 4.4 km s$^{-1}$. This component appears to be
distinct until $V_{\rm LSR}$ = 5.3 km s$^{-1}$, while it become merged in the
extended component at $V_{\rm LSR}$ = 5.7 -- 7.5 km s$^{-1}$.
At $V_{\rm LSR}$ = 8.0 -- 8.8 km s$^{-1}$ a gas component to the north of Source S
appears as a distinct gas component.
These blueshifted and redshifted model emission corresponds to the
observed component located between Source N and Source S,
and is also present in Figure \ref{fig:nspv},
while the amount of the velocity gradient in the model appears to be smaller
than that of the observed components (green lines in Figure \ref{fig:nspv}).
% (Figures \ref{fig:c18och}, \ref{fig:ocsch}, and \ref{fig:nspv}).
% The velocity structure of this emission component should reflect the gas motion inside the Roche sphere.
The blueshifted part of this component could be interpreted as the blueshifted counterpart of the CSD
around Source N, and the redshifted part the redshifted counterpart of the CSD around Source S.
Compared to the inferred Keplerian rotation profiles
of the individual CSDs (red and blue curves), the velocities of the observed and model components
between the protobinary are slower, suggesting slower azimuthal velocities.
Our model shows that a bridge gas structure connects the two CSDs, crossing
the L1 Lagrangian point \cite[see Figures 2 and 4 in][]{mat19}.
The gas flow has a stagnation point around the L1 Lagrangian point,
and it exhibits a linear velocity gradient in the P-V diagrams, which are consistent with
the observed velocity gradient as seen in the C$^{18}$O and OCS emission.
In particular, the velocity of the observed and model emission components located
between Source N and Source S at the L1 stagnation point
is consistent with the systemic velocity
(crossing point between the vertical black dashed lines and the L1 green dashed line).
% In other words, our ALMA observations have resolved the gas motion around the L1 point.

In Figure \ref{fig:nspv}, the observed high-velocity blueshifted gas components
to the south and north of Source N in the C$^{18}$O and OCS emission
are also seen. The model counterparts of these emission components are,
however, not present.
The velocity of the highest blueshifted component to the south of Source N
is too high as compared to the inferred Keplerian velocity of the CSD around Source N.
On the other hand, the projected location of the highly blueshifted component
to the north of Source N is almost identical to that of the redshifted emission
originated from the CSD rotation around Source N.
The high-velocity blueshifted components to the south and north of Source N
may thus be distinct gas components whose locations are above or below the disk plane.

% \begin{figure}[ht!]
% \figurenum{12}
% \epsscale{1.2}
% \plotone{c18opvmin_obsmod4v7.pdf}
% \caption{Observed and model P-V diagrams of the C$^{18}$O (3--2) emission
% along the cuts parallel to the minor axis of the CBD (P.A. = 70$\degr$) passing through
% $\pm$0$\farcs$75 offsets from the middle position between Sources N and S.
% The right panel shows the moment 0 map of the C$^{18}$O emission
% (contour) superimposed on the 0.9-mm dust continuum image (gray),
% as a guide map of the P-V cuts (dashed lines). In this guide map
% black crosses show the positions of the origins of the PVs.
% Contour levels in the P-V diagrams are from 3$\sigma$
% in steps of 2$\sigma$ (1$\sigma$ = 5.3 mJy beam$^{-1}$ = 4.5 K).
% Horizontal dashed lines denote the origin of the P-V diagrams.
% Green dashed lines delineate
% the detected velocity features. A vertical dashed line represents the
% systemic velocity of 6.4 km s$^{-1}$.
% \label{fig:c18opvmin}}
% \end{figure}

\subsection{Comparison between L1551 IRS 5 and NE; FU Ori CBD in L1551 IRS 5 ?} \label{subsec:fuori}

As well as the apparent absence of the infall motion in the CBD of L1551 IRS 5,
there are several intriguing differences between L1551 IRS 5 and NE,
both of which are Class I binary protostars
located within $\sim$2$\farcm$5 ($\sim$0.10 pc) in the L1551 region.
First, the outermost radius of the CBD in L1551 IRS 5 is $\sim$half that of
L1551 NE as seen in the 0.9-mm dust-continuum emission (Figure \ref{fig:neirs5}).
Our numerical models for both L1551 IRS 5 and NE take the different binary masses, mass ratios,
separations, and the centrifugal radii of the outer infalling envelopes into account
(see Table \ref{theory}).
The model of L1551 NE approximately reproduces the observed extent of the CBD \cite{tak14,tak17}.
On the other hand,
the extent of the model CBD in L1551 IRS 5 seems larger than
that of the observed CBD. In the model of L1551 IRS 5
the high temperature of the gas and
angular momentum transport by the gravitational torque of the binary stars
expands the CBD.
Thus, in the case of L1551 IRS 5 an additional mechanism to confine the CBD
is likely required, and a possible cause is the magnetic field.
A second difference is the spiral morphology.
In L1551 NE the western part of the CBD is much brighter
than the eastern part, suggesting the presence of the $m=1$ mode of the material distribution.
Such an $m=1$ mode in the spiral arms is less obvious in L1551 IRS 5.
Our model of L1551 IRS 5 adopts the high gas kinetic temperature of 100 K,
which suppresses the $m=1$ mode. On the other hand, we found that our models with lower temperatures
show the $m=1$ mode, even if the binary mass ratio $q$ is unity.
Thus, the difference of the $m=1$ mode between L1551 IRS 5 and NE
could be due to the difference of the gas kinetic temperature.

% This difference is reproduced with our numerical models.

Another unique difference between L1551 IRS 5 and NE is the difference of the
brightness temperatures in the
protobinary systems. In L1551 IRS 5, the peak brightness temperature of the 0.9-mm
dust-continuum emission in the CSD toward
Source N exceeds 260 K, and that toward Source S 160 K. By contrast,
in L1551 NE the peak brightness temperatures toward Source A and Source B are 42 K and 18 K,
respectively (Figure \ref{fig:neirs5}). In L1551 IRS 5, the peak brightness temperature
of the molecular lines in
the CBD exceeds 100 K (Figure \ref{fig:tb}), while that in L1551 NE is $\sim$54 K.
% in the $^{13}$CO (3--2) line.
These results imply that the kinetic temperature of the protobinary system L1551 IRS 5
is likely much higher than that of L1551 NE, as discussed above
regarding the $m=1$ mode.
The bolometric luminosity of L1551 IRS 5 is $L_{bol} \sim$22 $L_{\odot}$,
much brighter than that of L1551 NE ($L_{bol} \sim$4.2 $L_{\odot}$) \cite{fro05}.
On the other hand,
the total masses of the protostellar binaries as inferred from the CBD rotations are
$\sim$0.5 $M_{\odot}$ and $\sim$0.8 $M_{\odot}$ in L1551 IRS 5 and NE, respectively.
Thus, the central protostellar mass of the hotter system is lower.
These dynamical masses are below a solar mass,
and the bolometric luminosities are much higher than the stellar luminosities.
Thus, specific mass accretion burst in L1551 IRS 5
is likely required to account for the luminosity and temperature differences.

L1551 IRS 5 has been classified as an embedded FU Orionis object.
% from the presence of the first overtone CO absorption bands at 2.3 $\micron$ \cite{rei97}.
FU Orionis events are a brightening of the optical luminosities by 5--6 orders of magnitude
in months to years time scales, followed by the decay of the luminosity over several decades
to a century \cite{her77,har85,vor06}. The physical mechanism of the FU Orionis eruption is understood
as a sudden increase of the mass accretion in the disk onto the central stellar object.
Whereas the optical brightening of L1551 IRS 5 has not been
witnessed presumably because of the obscuration from the surrounding protostellar envelope,
the presence of the first overtone CO absorption bands at 2.3 $\micron$ is considered
as a piece of evidence of an FU Orionis event because the CO absorption arises from the cooler surface
layer of the hot accretion disk \cite{rei97}.
% L1551 IRS 5 is the brightest protostellar object in the entire Taurus star-forming region.
It is interesting to note that the CSD toward Source N
is hotter than that toward Source S by $\sim$100 K, even though the protostellar masses
of Source N and Source S are similar, % judging from the rotational motion in the CBD,
and that the western side of Source N is the hottest region in the CBD as seen
in the observed molecular lines (Figure \ref{fig:tb}). From these results,
we consider that Source N has experienced a more energetic accretion burst
than Source S, or that the accretion burst in Source N is the latest one
following that in Source S. The brighter central protostar Source N irradiates
the surrounding CSD, and through the cavity excavated by the blueshifted jet and outflow, 
the escaped radiation heats the surface of the CBD on the western side directly.
Because the western side is the far-side, there is less, cold obscuring material
along the line of sight, which makes the skewed temperature distribution (Figure \ref{fig:tb}).
The detected highly blueshifted components to the north and south of Source N,
whose origins are not understood with our numerical simulation, could be attributed
to the FU Orionis event.
% Since these emission components are strongly detected in the
% OCS emission at $E_u$ = 237 K as well as the C$^{18}$O emission, the gas components
% must be hot.
% This component may trace distinct infalling gas streams just landed onto Source N,
% which shows the high brightness temperature in the 0.9-mm dust-continuum emission.

Statistics of solar-type main-sequence binaries shows preference to
the equal mass ratio ($i.e.$, $q$=1),
while the distribution of the non-equal mass binaries is almost flat between $q \sim$0.1 and $< 1$
\cite{rag10,moe17,elb19}.
The binary mass ratio in L1551 IRS 5 is close to unity but that in L1551 NE is $q \sim$0.2.
Our simulations predict that the secondary
accretes more than the primary in L1551 NE \cite{tak14,tak17,mat19}.
The mass ratio of L1551 NE may thus approach to $q \sim$1.
On the other hand, our simulations with an equal binary mass predict
the equal mass accretion rate, which implies that the binary mass ratio
of L1551 IRS 5 will remain the same.
It is desirable to observationally measure the mass accretion rates
onto the primary and secondary of L1551 IRS 5 and NE, as well as
the other protobinary sources, to understand the origin of the distribution
of the binary mass ratio. One of the key points is to resolve well
the structures and gas motions inside the Hill radii, and to compare
the observational results to those of the numerical simulations.
These should be the subject to our next researches.

% Unfortunately, as discussed in the last subsection
% it is not straightforward to observationally identify infall toward
% the individual binary protostars and to measure the mass accretion rates.
% It is necessary to construct computationally less-expensive CBD models
% and to perform fitting of the models to the observational image cubes,
% in order to derive the mass accretion rates.

\section{Summary} \label{sec:sum}

We have conducted ALMA Cycle 4 observations of the protostellar binary system
L1551 IRS 5 in the 0.9-mm dust-continuum emission,
and the C$^{18}$O (3--2), $^{13}$CO (3--2), CS (7--6), OCS (28--27),
SO (7$_8$--6$_7$), and HC$^{18}$O$^{+}$ (4--3) lines
at an angular resolution of $\lesssim$0$\farcs$1.
We have also performed hydrodynamical simulations, radiative transfer calculations,
and ALMA observing simulations to make theoretically-predicted 0.9-mm and C$^{18}$O images of L1551 IRS 5
and to interpret the observed features.
The results of L1551 IRS 5 are compared to our ALMA Cycle 2 results of L1551 NE,
other Class I protostellar binary located just $\sim$2$\farcm$5 ($\sim$0.10 pc) northeast of L1551 IRS 5.
The main results are summarized below.

\begin{itemize}
\item[1.] The 0.9-mm image resolves two circumstellar disks (CSDs) around the individual binary
protostars (Source N and Source S) and the circumbinary disk (CBD) in L1551 IRS 5.
The relative location of the binary protostars as observed in 2017 is consistent with
the identified clockwise orbital motion starting from 1983.
The beam-deconvolved size of the CSDs
is $\sim$20 au, and the outermost radius of the CBD $\sim$150 au.
The ringlike CBD feature that previous ALMA observations found has been
resolved into two spiral arms with the present observations,
one arm connecting to the CSD in Source N and the other to Source S.
The 0.9-mm peak brightness temperatures of the CSDs
toward Source N and Source S reach 260 K and 160 K, respectively, and the spectral index $\alpha$ between 0.9 mm and 1.3 mm
in the CSDs is as low as 2. These results imply that the 0.9-mm continuum emission
from the CSDs is optically thick and traces the dust temperature.
The peak brightness temperature of the 0.9-mm dust-continuum emission toward the CBD
is $\lesssim$20 K and the $\alpha$ value $\sim$3 -- 3.7, suggesting that the 0.9-mm emission from the CBD is
optically thin. The mass of the CBD is estimated to be 0.015--0.060 $M_{\odot}$ for $T_d$ = 30--100 K.
As compared to the CBD image at 0.9-mm in L1551 NE, the CBD in L1551 IRS 5 is half the size
but the brightness temperature is much higher. Furthermore, the $m$ = 1 mode of the two-arm spirals
found in L1551 NE is less obvious in L1551 IRS 5.

\item[2.] All the observed molecular lines except for the HC$^{18}$O$^{+}$ (4--3) line have been
strongly detected toward the CBD.
% plus signs of contamination of the associated molecular outflows
In particular, the OCS (28--27) emission with
the upper-state energy of the rotational level of 237 K
has been detected. % in the inner part of the CBD.
% as seen in the 0.9-mm emission.
The peak brightness temperatures of the detected molecular lines are as high as $\gtrsim$100 K.
These results imply the high gas kinetic temperature of the CBD.
The western side of Source N in the CBD, which corresponds to the far-side, is the brightest,
where the intense radiation from Source N as inferred from the high brightness temperature
of the CSD in the 0.9-mm continuum emission is reflected.
Because of the intense background continuum emission,
the molecular lines show negative intensities toward the CSDs.

\item[3.] The velocity channel maps of the C$^{18}$O and OCS emission show
distinct, multiple velocity components. Primary emission components
are a blueshifted (3.5 -- 5.3 km s$^{-1}$) component to the southeast and
a redshifted (7.1 -- 9.3 km s$^{-1}$) component to the northwest of the protobinary,
which originates from the rotation of the CBD.
Both of these blueshifted and redshifted components also exhibit velocity gradients
along the northeast (blueshifted) - southwest (redshifted) orientation.
These results can be interpreted as the expanding motion of the rotating CBD.
% caused by the faster rotation than the local Keplerian rotation of the CBD.
In the highly blueshifted
($V_{LSR}$ = 1.9 - 3.1 km s$^{-1}$) and redshifted (9.7 - 11.3 km s$^{-1}$)
velocities, compact molecular emission are located to the south
of Source S and north of Source N, respectively. These components
are likely originated from the rotating motions of the individual CSDs,
and exhibit transitions from the CBD to the CSD rotations at around
the L2 and L3 Lagrangian points.
A gas component located between Source N and Source S with
a north (blueshifted) to south (redshifted) velocity gradient
is also observed at $V_{LSR}$ = 3.1 - 8.7 km s$^{-1}$.
This component should reflect the gas motion around the L1 point.
There are another, highly blueshifted components (-2.7 -- -0.3 km s$^{-1}$
and 1.3 -- 3.6 km s$^{-1}$) to the south and north of Source N, respectively.

\item[4.] Our numerical model reproduces the observed two spiral-arm structure as seen
in the 0.9-mm dust-continuum image, whereas the extent of the model CBD is larger than
that observed. The observed confinement of the CBD may be attributed to the effect
of the magnetic field, which is not included in our numerical model.
Our model velocity channel maps of the C$^{18}$O emission reproduce the observed expanding gas motion,
due to the faster rotation than the Keplerian rotation caused by
the non-axisymmetric gravitational torque of the binary in the spiral regions.
Infall in the inter-arm regions predicted in our numerical model is not observationally identified,
because the projected locations of the binary are closely aligned to the major axis
of the CBD and the infall regions reside along the major axis.
The transitions from the CBD to the CSDs at the L2 and L3 Lagrangian points are
also reproduced with our model.
Furthermore, our numerical model demonstrates that the observed gas component located between
the protobinary with the north (blueshifted) - south (redshifted) velocity gradient can be interpreted 
as the gas flow around the stagnation point, which is located at the L1 Lagrangian point.
On the other hand, the model velocity channel maps do not reproduce the observed 
highly blueshifted components to the north and south of Source N.

\item[5.] The brightness temperatures of the 0.9-mm + molecular emission in the CSDs and CBD
and the bolometric luminosity are much higher in L1551 IRS 5 than those in L1551 NE,
whereas the inferred protobinary mass of L1551 IRS 5 ($\sim$0.5 $M_{\odot}$) is lower
than that of L1551 NE ($\sim$0.8 $M_{\odot}$). These results
support the previous claim that L1551 IRS 5 is in the state of an FU Orionis event.
The absence of the $m$=1 mode of the spirals in L1551 IRS 5, in contrast to that in L1551 NE,
can be attributed to the high gas temperature in the CBD.
The observed highly blueshifted components around Source N, which exhibits a higher brightness
temperature in the CSD than Source S, may be related to the FU Orionis event.
\end{itemize}

%% The "ht!" tells LaTeX to put the figure "here" first, at the "top" next
%% and to override the normal way of calculating a float position

\acknowledgments
We are grateful to the anonymous referee for insightful comments and detailed reading of the manuscript.
We would like to thank all the ALMA staff supporting this work.
S.T. and P.T.P.H. acknowledge grants from the Ministry of Science and Technology (MOST) of Taiwan,
MOST 102-2119-M-001-012-MY3 and 108-2112-M-001-016-MY1, respectively.
S.T., T.M., and M.S. are supported by
JSPS KAKENHI grant Nos. 16H07086 and 18K03703,
17H02863, 17K05394, 18H05436, and 18H05437,
and 16K05303, respectively.
% JSPS KAKENHI grant Nos. 16H07086 and 18K03703 in support of this work.
% T.M. is supported by the KAKENHI grants Nos. 17H02863, 17K05394,
% 18H05436, and 18H05437, and M.S. is supported by 
J.L. is supported by the GRF grants of the Government of the Hong Kong SAR under HKU 703512P.
L.W.L. acknowledges support from NSF AST-1910364.
This work was supported by NAOJ ALMA Scientific Research grant No. 2017-04A.
This paper makes use of the following ALMA data:
ADS/JAO.ALMA\#2016.1.00138.S
ALMA is a partnership of ESO (representing
its member states), NSF (USA) and NINS (Japan), together with NRC (Canada)
and NSC and ASIAA (Taiwan) and KASI (Republic of Korea), in
cooperation with the Republic of Chile. The Joint ALMA Observatory is
operated by ESO, AUI/NRAO and NAOJ.
Numerical computations were in part carried out on Cray XC50 at
Center for Computational Astrophysics, National Astronomical Observatory of Japan.


%% for the paper.  Each keyword is check against the master list during
%% copy editing.  Individual instruments can be provided in parentheses,
%% after the keyword, but they are not verified.

\vspace{5mm}
\facilities{ALMA}
\software{CASA \cite{mcm07}, Miriad \cite{sau95}}

\listofchanges

\end{document}